\documentclass[]{article}
\usepackage[margin=1in]{geometry}

\usepackage{lipsum}
\usepackage{amsfonts}

\usepackage{float}
\usepackage{flafter}
\usepackage[dvips]{epsfig}
\usepackage{mathtools}
\usepackage{xcolor}
\usepackage{dsfont}
\usepackage{tikz}
\usetikzlibrary{automata,positioning}
\usetikzlibrary{shapes.geometric, arrows}
\usetikzlibrary{arrows.meta}
\usepackage{pict2e}
\usepackage{amsmath}
\usepackage[utf8]{inputenc}
\usepackage[english]{babel}
\usepackage{graphicx}
\usepackage{comment}
\usepackage{epic}
\usepackage{lscape}
\usepackage{rotating}
\usepackage{epstopdf}
\usepackage{wrapfig}
\usepackage{amssymb}
\usepackage{bm}
\usepackage{subcaption}

\ifpdf
  \DeclareGraphicsExtensions{.eps,.pdf,.png,.jpg}
\else
  \DeclareGraphicsExtensions{.eps}
\fi

\title{Nonlinear  shallow-water waves with vertical odd viscosity}

\author{Alex Doak\thanks{Corresponding author. Department of Mathematical Sciences, University of Bath, UK BA2 2AY. {\color{blue}add49@bath.ac.uk}}
	\and Guido Baardink\thanks{Department of Physics, University of Bath, UK BA2 2AY.}
	\and Paul A Milewski \thanks{ Department of Mathematical Sciences, University of Bath, UK BA2 2AY}%\footnotemark[2]
	\and Anton Souslov\footnotemark[3] \thanks{A.S.~acknowledges the support of the Engineering and Physical Sciences Research Council (EPSRC) through New Investigator Award No.~EP/T000961/1 and of the Royal Society under grant No.~RGS/R2/202135.}
}

\usepackage{amsopn}

\DeclareMathOperator{\sech}{sech}
\usepackage{booktabs}% http://ctan.org/pkg/booktabs

\usepackage{array}
\usepackage{overpic}

\usepackage{tabularx}
\usepackage{wasysym}

\newcommand{\Rey}
{%
	\ensuremath
	{
		\nu %\left(\textrm{Re}^{o}\right)^{-1}
	}
}

\newcommand\h[1]
{%
	\ensuremath
	{
		^{(#1)}
	}
}

\newcommand\lrr[1]
{%
	\ensuremath
	{
		\left( #1 \right)
	}
}
\newcommand\lrs[1]
{%
	\ensuremath
	{
		\left[ #1 \right]
	}
}

\newcommand\e
{%
	\ensuremath
	{
		\epsilon
	}
}

\newcommand\uu[1]
{%
	\ensuremath
	{
	{#1}	%\underline{\underline{#1}}
	}
}

\newcommand\ol[1]
{%
	\ensuremath
	{
		\overline{#1}
	}
}

\usepackage{soul}

\date{May 10, 2023}

\usepackage{pgfplots}
\usepackage[numbers]{natbib}
\tikzstyle{box1} = [rectangle, rounded corners, minimum width=3cm, minimum height=1cm,text centered, draw=black, fill=red!30]
\tikzstyle{box2} = [rectangle, rounded corners, minimum width=3cm, minimum height=1cm,text centered, draw=black, fill=green!30]
\tikzstyle{box3} = [rectangle, rounded corners, minimum width=3cm, minimum height=1cm,text centered, draw=black, fill=yellow!30]

\usepackage[normalem]{ulem}
\begin{document}

\maketitle

\begin{abstract}
The breaking of detailed balance in fluids through Coriolis forces or odd-viscous stresses has profound effects on the dynamics of surface waves.
Here we explore both weakly and strongly non-linear waves in a three-dimensional fluid with vertical odd viscosity with and without the Coriolis effect.
Our model describes the free surface of a shallow fluid composed of nearly vertical vortex filaments, which all stand perpendicular to the surface. We find that the odd viscosity in this configuration induces previously unexplored non-linear effects in shallow-water waves, arising from both stresses on the surface and stress gradients in the bulk. By assuming weak nonlinearity, we find reduced equations including Korteweg-de Vries (KdV), Ostrovsky, and Kadomtsev-Petviashvilli (KP) equations with modified coefficients. At sufficiently large odd viscosity, the dispersion changes sign, allowing for compact two-dimensional solitary waves. We show that odd viscosity and surface tension have the same effect on the free surface, but distinct signatures in the fluid flow.
Our results describe the collective dynamics of many-vortex systems, which can also occur in oceanic and atmospheric geophysics.
\end{abstract}

\section{Introduction}

Fluids subject to internal rotations  can acquire  interesting mechanical properties, which only recently have started to be explored. For example, these fluids have in common  the so-called breaking of detailed balance (that is, the absence of microscopic reversibility away from equilibrium),   leading to the breakdown of Onsager reciprocal relations, which would otherwise enforce a symmetric viscosity tensor. 
As a consequence the general viscosity tensor can acquire new components prohibited in typical fluids (e.g., Newtonian fluids) leading to dispersive rather than dissipative effects. These new viscosity coefficients have been collectively termed \emph{odd viscosity}~\cite{avron1998odd} (equivalently, Hall viscosity~\cite{avron1995viscosity}).
Both Coriolis forces and internal rotation violate detailed balance: in the case of Coriolis
forces, this violation arises from a non-inertial frame of reference, whereas in the internal rotation case, the violation arises from the coarse-graining of the effect of spinning fluid particles. These flows are also chiral: they have a handedness induced by the sign of the rotation of the frame of reference or of the internal spin~\cite{banerjee2017odd}. %\blue{... include reference [8]?}

%Odd viscosity has a broad range of theoretical and experimental implications.
Odd viscosity occurs in a variety of physical systems. For example, interactions between vortices, or more generally, spinning constituents, have a transverse character, leading to the characteristic phenomenology of so-called vortex fluids. The term vortex fluid is broadly applied to describe a coarse-grained fluid composed of rotating constituents, which have a single characterstic length and time scale. For example, quantized vortices arise in response to global rotation in superfluid helium ~\cite{avron1995viscosity, berdyugin2019measuring}, and chiral active fluids support long-lived vortex states due  to external injection of angular momentum, for example using particles that are rotated with an external field \cite{banerjee2017odd,bililign2021motile,han2021fluctuating,soni2019}. As a simple model of transient vortex-fluid states, Refs.~\cite{bogatskiy2019edge,Wiegmann2014} considered a fluid flow induced by a distribution of two-dimensional point vortices in an inviscid fluid. In all these cases, odd viscosity is an emergent novel behaviour arising due to the simultaneous chirality and breaking of detailed balance.

In two dimensions, odd viscosity can remain isotropic, for example when the particles all have rotation normal to the plane of the flow, or anisotropic~\cite{gromov2017investigating,Offertaler2019,souslov2020anisotropic, rao2020hall}. By contrast, in three dimensions, odd viscosity must be anisotropic~\cite{avron1998odd,robredo2021cubic}.
The hydrodynamic consequences of odd viscosity in three-dimensional fluids has received less attention.
For example, Ref.~\cite{markovich2021odd} derived odd viscosity for a three-dimensional incompressible fluid from a Hamiltonian model of dissipationless spinning particles. In the opposite limit of Stokes flow dominated by dissipation, odd viscosity creates parity-violating flows under conditions as common as sedimentation~\cite{khain2020stokes}. 
For both sound~\cite{baardink2021complete,souslov2019topological} and linear gravity waves~\cite{Tauber2019,Tauber2020}, odd viscosity leads to topological boundary modes.

\begin{figure}
    \centering
    \includegraphics[width=.8\linewidth]{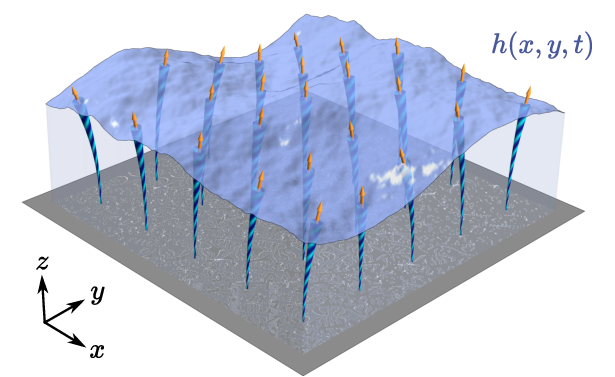}
    \caption{Schematic of the model considered. The flow is bounded below by a flat wall at $z=0$, and above by a free surface $z=h(x,y,t)$. The fluid is composed of a distribution of vortex filaments, which remain perpendicular to both boundaries.
    }
    \label{fig:setup}
\end{figure}

Odd viscosity has complex and profound effects on the behavior of surface waves even in two dimensions, where the free surface is a one-dimensional curve. Experimentally, odd viscosity has been measured through its effects on the linear dispersion of these waves~\cite{soni2019}.  In a boundary layer at the fluid surface, odd viscosity can lead to effects akin to a surface tension, but with broken detailed balance~\cite{abanov2018odd, abanov2019free}. 
In the nonlinear regime, these boundary layers interact with capillary effects~\cite{granero2021motion} or   compressibility~\cite{abanov2020hydrodynamics}, and modify the coefficients of the Korteweg-de Vries (KdV) equation in shallow water~\cite{monteiro2020non}. 
In all of these two-dimensional cases, odd viscosity has been assumed pointing out-of-plane and tangentially to the surface. This geometry occurs experimentally, for example, when self-rotating particles in a layer spin around the axis which is out of plane~\cite{soni2019}. By contrast, we focus on nonlinear surface waves in three-dimensional geometries in which odd viscosity arises from rotations that point normal to the surface. Our geometry with vertical odd viscosity occurs, for example, for a free surface above a vortex fluid, see Fig.~\ref{fig:setup}. 
Our model describes a minimal and generic three-dimensional vortex fluid, and  captures the effects of odd viscosity on the nonlinear propagation of surface waves.

 In contrast to odd viscosity, the effect of Coriolis forces on surface waves has been extensively explored due to its importance in geophysics (see ~\cite{cushman2011introduction,pedlosky1987geophysical}, and references therein). Coriolis forces result from the rotation of the Earth or, more generally, from considering waves in a rotating frame of reference. Both Coriolis and odd-viscous terms break detailed balance, but Coriolis forces also violate invariance under change of inertial reference frame, i.e., Galilean invariance. 
 Although vortices are prevalent in planetary oceans and atmospheres, the potential geophysical consequences of resulting odd-viscous stresses remain unexplored. In this paper, we explore the effects of both odd viscosity and Coriolis forces (i.e., a fluid subject to both internal and external rotations).
 
 For nonlinear surface waves, a common starting point is the shallow-water approximation. In this approximation, the three-dimensional fluid is described by the dynamics of its two-dimensional free surface and the averaged horizontal velocities, assuming the depth of the fluid to be much smaller than the typical surface wavelength. Leading order dispersive (i.e., non-hydrostatic) effects can be added while keeping the system strongly nonlinear. For gravity waves, the resulting equations (without odd viscosity) were first derived by Serre \cite{serre1953contribution} and Green-Naghdi \cite{green1976derivation}, and subsequently extensively explored~\cite{dias2010fully,jalali2021balance}. They have been shown to accurately represent solutions to the Euler equations and compare well with experiments \cite{dutykh2013finite,camassa2006realm}.
 
 In this paper, we derive the non-linear Serre equations with odd viscosity and the Coriolis force, which model the flow of a 3D fluid composed of many vertical vortices, bounded above by a free surface (Fig.~\ref{fig:setup} and Section 2). We 
 average quantities across the depth of the fluid
 and use the shallow-water approximation to obtain two-dimensional non-linear equations describing the evolution of the fluid velocity and the free surface (Section 3). We then use a hierarchy of weakly nonlinear approximations (see Fig.~\ref{fig:overview} and Section 4) to find analytical solutions and compare them with numerical solutions to the odd-viscous Serre equations (Section 5). 
 
While we derive the general evolutionary equations for the wave motion, we focus particularly on their solitary-wave solutions. Consistent with intuition, both odd viscosity and Coriolis forces induce flows perpendicular to the propagation direction of a planar solitary wave. In the Serre equations, odd viscosity induces new non-linear terms proportional to the stress tensor itself. In the reduced equations, which include the one-dimensional Korteweg-de Vries (KdV), two-dimensional Kadomtsev-Petviashvili  (KP,~\cite{kadomtsev1970stability}), and the rotating Ostrovsky-type equations~\cite{ostrovsky1978nonlinear, grimshaw1989derivation}, odd viscosity only enters as a parameter modifying the dispersion. At sufficiently large odd viscosity, the dispersion changes sign, allowing for localised two-dimensional solitary waves in the KP equation. We show that, in the weakly nonlinear regimes, odd viscosity and surface tension lead to identical free-surface shapes. However, unlike surface tension, odd viscosity breaks both detailed balance and chirality, leading to distinct fluid flows with a transverse component.

\begin{figure}
\centering
\begin{tikzpicture}[scale=1, every node/.style={scale=0.8}]
\node (in0) [box1, text width=3cm] {3D rotating Euler equations};
\node (in1) [box1, below of=in0, yshift=-1cm, text width=3cm] {3D rotating Euler with \textbf{odd viscosity},  Eqs.~(\ref{eq:cont}--\ref{eq:momxy}) };
\node (in2) [box2, below right of=in1, xshift=4cm, yshift=-1.5cm, text width=3cm] {2D Serre with \textbf{odd viscosity}, Eqs.~(\ref{eq:RSWE1}--\ref{eq:RSWEcon2})};
\node (in3) [box2, below of=in2, yshift=-3cm, text width=3cm] {2D Rotation-modified KP, Eq.~\eqref{eq:RMKP}};
\node (in41) [box2, below of=in3, yshift=-1.2cm, text width=3cm] {2D KP, \\ Eq.~\eqref{eq:KP}};
\node (in42) [box3,  right of=in3, xshift=4.5cm, text width=2.5cm] {1D Ostrovsky, Eq.~\eqref{eq:OS}};
\node (in5) [box3, below of=in42,yshift=-1.2cm, text width=2.5cm] {1D KdV, Eq.~\eqref{eq:KdV}};
\node (in6) [box3, right of=in2,  xshift=4.5cm,text width=2.5cm] {1D NLS, Eq.~\eqref{eq:NLS}};
\draw [shorten >=0.1cm,shorten <=0.1cm, ->] (in0) -- node[anchor=west] {Distribution of vortices}  (in1);
\draw [shorten >=0.2cm,shorten <=.2cm, ->] (in1) -- node[anchor=west, align=left, yshift=.5cm] {Shallow water approximation, $\mu \ll 1$  \\ Depth averaging } (in2);
\draw [shorten >=0.1cm,shorten <=0.1cm, ->] (in2) -- node[anchor=north,yshift=-.1cm, xshift=0.1cm,text width=3cm] {\begin{center}Quasi-monochromatic planar wave\end{center}}  (in6);
\draw [shorten >=0.2cm,shorten <=.2cm, ->] (in2) -- node[anchor=east] {} node[anchor=east,align=right] {Uni-directional weakly nonlinear, $h=1+\epsilon \eta$ \\ Weak rotation,  $f=\sqrt{\epsilon} \hat{f}$ \\  Weak $y$ variance, $y=\sqrt{\epsilon}\hat{y}$ \\
Boussinesq scaling, $\epsilon\sim\mu^2$, $T=\epsilon t$ }
(in3);
\draw [shorten >=0.2cm,shorten <=.2cm, ->, red] (in3) -- node[anchor=center,left,xshift=-.1cm, purple] {No rotation, $\hat{f}=0$} (in41);
\draw [shorten >=0.2cm,shorten <=.2cm, ->, blue] (in3) -- node[anchor=north, blue,yshift=-.1cm] {Plane waves} (in42);
\draw [shorten >=0.2cm,shorten <=.2cm, ->, blue] (in41) --  node[anchor=north, blue,yshift=-.1cm] {Plane waves} (in5);
\draw [shorten >=0.2cm,shorten <=.2cm, ->, red] (in42) -- node[anchor=west,xshift=.1cm, purple] {No rotation, $\hat{f}=0$} (in5);
\end{tikzpicture}
\caption{Overview of the equations seen in the paper. The system is simplified via additional assumptions as one moves along the arrows. The red, green, and yellow boxes represent 3D, 2D, and 1D systems of equations, respectively. \label{fig:overview}}
\end{figure}

\section{Formulation}\label{section:ref}
In this paper, we will explore shallow-water nonlinear theories to model free-surface flows in an incompressible vortex fluid of constant density $\rho$, for which the classical viscous dissipation term is taken to be negligible. The vortex fluid contains a distribution of almost vertical vortices at the smallest, microscopic scales. The fluid is bounded below by a flat bottom and above by a free surface. The effect of the vorticity is assumed to enter the equations of motion via a modification to the classical Cauchy stress tensor, arising from a coarse graining of the point vortices, and resulting in what is known in literature as odd viscosity \cite{avron1998odd}.

We follow the formulation for the problem and the coarse-graining approximation of Ref.~\cite{Wiegmann2014}, where an effective Euler equation for a two-dimensional point-vortex flow is derived for a vortex velocity and a vortex density. The vortex density is materially conserved and the vortex velocity satisfies a momentum equation with a dispersive correction arising from the odd viscosity. We consider the simpler case where the vortex density is constant when the free surface is undisturbed. Then, as a consequence of the shallow-water conservation of potential vorticity, the vertical vorticity density as defined by Ref.~\cite{Wiegmann2014} is conserved, leaving only changes in the momentum equation to be considered. Throughout this paper, horizontal velocities denote the coarse-grained vortex velocities.

 We consider Cartesian coordinates $(x,y,z)$, and denote the velocities in the $x$, $y$, and $z$ direction as $u$, $v$, and $w$, respectively. The surface of the fluid is denoted as $z=h(x,y,t)$, and the fluid has an undisturbed depth of $H$. We choose the wall bounding the fluid from below to be at $z=0$. Gravity acts in the negative $z$-direction, perpendicular to the undisturbed interface and the rotation is about the $z$-axis. The flow configuration is shown in Figure~\ref{fig:setup}. For generality, we will consider the fluid to be in a rotating reference frame and include the Coriolis effect. Throughout the paper, we find it helpful to separate motion in the $xy$-plane and in the $z$-direction.  For this purpose, we introduce the vector $\bm{u}=[u,v]$, and the operator $\nabla_x=[\partial_x,\partial_y]$. For compactness, we also introduce the vector $\bm{v}=[u,v,w]$ and operator $\nabla=[\partial_x,\partial_y,\partial_z]$. The three-dimensional equations of motion are given by
\begin{align}
\nabla \cdot \bm{v} &=0, \label{eq:cont}\\
    \rho \bm{v}_t +  \rho \lrr{\bm{v}\cdot\nabla} \bm{v} &=   \nabla \cdot \uu{\sigma}
    +  \beta \rho  \bm{v}^*, \label{eq:momxy}
\end{align}
Here, $\bm{v}^*=[\bm{u}^*,0]$ and $\bm{u}^*\equiv\epsilon_{ij}\bm{u}_j = [v, -u]$, where $\epsilon_{ij}$ is the two-dimensional Levi-Civita symbol, $\beta$ is the Coriolis coefficient, and $\sigma$ is the Cauchy stress tensor, given by
\begin{align}\label{eq:sigma}
\uu{\sigma} &=  -p \uu{I_d}
+\rho \nu^o  \uu{T}.
\end{align}
Note that $p$ is the pressure variation from hydrostatic pressure. One can recover the absolute pressure $p_a$ via the equation $p_a=p-\rho g z$. The tensor $\uu{I_d}$ is the identity matrix, while $\uu{T}$ is associated with the distribution of vortex filaments. The constant $\nu^o$ is the kinematic odd viscosity, related to the vortex density and strength. We assume that the microscopic vortex filaments induce additional stresses which appear primarily as an odd viscosity in the $xy$-plane. The tensor $\uu{T}$ captures the effects of odd viscosity, and can be decomposed in orders of the shallow water parameter $\mu= H/L$ (see Section~\ref{section:nd}). This is discussed further in appendix $A$, where we denote the leading order contribution $K$.  The term $\uu{K}$ corresponds to the contribution induced by purely vertical filaments, which is given by
\begin{align}\label{eq:K}
\uu{T} &= \mu K + O(\mu^2), &&
\uu{K} =
    \begin{bmatrix}
 u_y+v_x & v_y - u_x & 0 \\
 v_y-u_x & -( v_x+u_y) & 0 \\
0&0&0
    \end{bmatrix}.
\end{align}
The upper left $2\times2$ submatrix of $K$ can be compactly expressed as a linear combination of strain-rate components, \begin{equation}
K_{ij} = \nabla_i^* \bm{u}_j + \nabla_i \bm{u}^*_j.
\end{equation}
where $\nabla_i^*\equiv\epsilon_{ij}\nabla_j$.
It is shown in Appendix~\ref{appendix:OV} that, when nondimensionalised, the strain rates $K$ are identical to the rates $T$ up to leading order in $\mu$ (see equation~\eqref{eq:rot2}).
There are small correction terms in the relationship between $T$ and $K$, attributed to the bending of the vortex filaments such that they remain perpendicular to the lower and upper boundaries.

Kinematic boundary conditions at the bottom wall and the free surface are given by
\begin{align}
w&=0, && \text{at} \,\,\, z=0, \label{eq:wallkbc}\\
w&=h_t + \bm{u}\cdot \nabla_x h, && \text{at} \,\,\, z=h(x,y,t) \label{eq:fskbc}
\end{align}
Finally, the dynamical boundary condition on the free surface is given by
\begin{align}\label{eq:fsdbc0}
\sigma_{ij} n_j&= -\rho g h n_i, && \text{at} \,\,\, z=h(x,y,t), 
\end{align}
where $\bm{n}$ is the unit normal to the free surface.
Noting that $T_{ij}n_j=0$ at the surface (see Appendix~\ref{appendix:OV}), this reduces to
\begin{align}
    p &=\rho g h - \gamma \nabla_x \cdot \lrr{\frac{\nabla_x h}{\sqrt{1+\lvert \nabla_x h \rvert ^2}}}, && \text{at} \,\,\, z=h(x,y,t), \label{eq:fsdbc}
\end{align}
where $\gamma$ is the surface tension coefficient, responsible for a pressure jump proportional to the mean curvature as given by the Young-Laplace equation.
%For simplicity we have not yet included the effect of surface tension. In later sections, we \blue{contrast the effects of odd viscosity and surface tension. For a derivation of the two-dimensional Serre equations with surface tension, see \cite{khorbatly2018derivation}. } \sout{bring back the surface tension to contrast its effects with those of odd viscosity.}.

\subsection{Nondimensionalisation and scaling}\label{section:nd}
We nondimensionalise the equations using the depth of the fluid $H$ as the vertical length scale and denote the horizontal length scale $L$.
The shallow water parameter is given by $\mu=H/L$, and since we are exploring shallow-water theory we assume $\mu\ll 1$. We denote by $U=\sqrt{gH}$ a typical velocity scale in the $xy$-plane, which we take as a reference velocity. This implies a time scale of $L/U$, and a pressure scale of $\rho U^2$. It follows from the incompressibility condition~\eqref{eq:cont} that the velocity in the $z$-direction is of dimension $\mu U$. We write
\begin{align}
(x,y)&=L(\tilde{x},\tilde{y}),&& z=H\tilde{z} && t=\frac{L}{U}\tilde{t}, && \bm{u}=U\tilde{\bm{u}}, && w=\mu U \tilde{w}, && p=\rho U^2 \tilde{p},
\end{align}
where tildes denote dimensionless variables. 
 In dimensionless form, upon dropping tildes, the system~\eqref{eq:cont}-\eqref{eq:momxy} becomes
\begin{align}
\nabla_x \cdot \bm{u} + w_z &=0, \label{eq:ndcont}\\
\bm{u}_t + \lrr{\bm{u}\cdot\nabla_x}\bm{u} + w \bm{u}_z & = -\nabla_x p + f \bm{u}^* \nonumber \\
+ \Rey \mu & \lrr{ \nabla_x \cdot \begin{bmatrix} T_{11}&T_{12}\\T_{21}&T_{22} \end{bmatrix} + \partial_z\begin{bmatrix} T_{31}, T_{32}    \end{bmatrix}^T}, \label{eq:ndmomxy}\\
 \mu^2 \lrr{ w_t + \lrr{\bm{u}\cdot\nabla_x}w + w w_z} + p_z &=  \Rey \mu^3  \lrr{\nabla \cdot \begin{bmatrix} T_{13}, T_{23} , T_{33}    \end{bmatrix}^T} . \label{eq:ndmomz}
\end{align}
We have split the momentum equations into two horizontal momentum equations, given by~\eqref{eq:ndmomxy}, and one vertical momentum equation~\eqref{eq:ndmomz}. The orders of the stress tensor components $T_{ij}$ are obtained from equation~\eqref{eq:ap_stress} in the Appendix, which followed assumptions about the form of $\uu{T}$. 
Two non-dimensional constants $\nu$ and $f$ arise, given by
\begin{align}
\Rey&= \frac{\nu^o}{UL} , && f=\frac{L\beta}{U}.
\end{align}
The nondimensional constant $\Rey$ is the inverse odd Reynolds number, which is a ratio of inertia and odd-viscous stresses as used in \cite{banerjee2017odd}, while $f$ is the Rossby number, the ratio of Coriolis to inertial effects. 
The boundary conditions are unchanged except for~\eqref{eq:fsdbc} which becomes
\begin{align}\label{eq:ndfsdbc}
p  &=  h - \mu^2 B \Delta_x h + O(\mu^4)  , && \text{at} \,\,\,\, z=h(x,y,t),
\end{align}
where $\Delta_x=\nabla_x \cdot \nabla_x$ and $B=\gamma/g\rho H^2$ is the nondimensional Bond number.
In the following section, we apply the shallow water approximation, and along with introducing depth averaged quantities, derive a nonlinear long-wave approximation to the above system. 

\section{Shallow-water approximation and depth averaging}\label{section:serre}
In this section, we will simplify the system of equations~\eqref{eq:ndcont}-\eqref{eq:ndmomz} by both truncating the model to order $O(\mu^3)$, and by exploiting depth averaged quantities. These simplifications results in a reduction in the dimensionality of the system.
Consider first the horizontal momentum equation~\eqref{eq:ndmomxy}. Substituting in the stress tensor~\eqref{eq:Tfinal}, the odd-viscous component of the equation is given by
\begin{align}
 \mu \Rey \lrr{ \nabla_x \cdot \begin{bmatrix} K_{11}&K_{12}\\K_{21}&K_{22} \end{bmatrix} + \partial_z\begin{bmatrix} \lrr{h_x K_{11} + h_y K_{12}}q \\ \lrr{h_x K_{21} + h_y K_{22}}q   \end{bmatrix}} + O(\mu^3)  \label{eq:ndmomxy2}.
\end{align}
Here, $q=z/h+O(\mu)$ is an interpolating function (see equation~\eqref{eq:g} in the Appendix). It follows from equations~\eqref{eq:ndmomxy2} that if $\bm{u}_z=O(\mu^2)$ at $t=0$, all terms with $z$-dependence occurring in the horizontal momentum equation~\eqref{eq:ndmomxy} will occur at $O(\mu^2)$. In other words, assuming the flow initially satisfies $\bm{u}_z=O(\mu^2)$ at $t=0$, it will do for all time. Imposing this condition, we write
\begin{align}\label{eq:expansion}
    \bm{u}(x,y,z,t) &= \overline{\bm{u}}(x,y,t) + \mu^2 \bm{u}^{(2)}(x,y,z,t) + O(\mu^3), \\
    p(x,y,z,t) &= p^{(0)}(x,y,t) + \mu^2 p^{(2)}(x,y,z,t) + O(\mu^3) ,
\end{align}
where we have introduced a depth-averaging operator on fluid variables, defined by
\begin{align}
\overline{A}(\bm{x},t) &= \frac{1}{h(\bm{x},t)} \int_{0}^h A(\bm{x},z,t) \, \mathrm{d} z.
\end{align}
From the above, it follows that $h \overline{A_s} = (h \overline{A})_s - A(\bm{x},h,t) h_s$, where $s$ is any independent variable and $h \overline{A_z} = A(\bm{x},h,t) - A(\bm{x},0,t)$.

Averaging the incompressibility condition~\eqref{eq:ndcont}, and making use of the kinematic boundary conditions~\eqref{eq:wallkbc}-\eqref{eq:fskbc}, we obtain the exact conservation of mass equation
\begin{equation}\label{eq:contda}
    h_t + \nabla_x \cdot\lrr{h \overline{\bm{u}}} =0.
\end{equation}
Our goal is to find a system of equations for $\bar{\bm{u}}$ and $h$. 
Next, therefore, we derive depth-averaged momentum equations by first re-writing the horizontal momentum equation~\eqref{eq:ndmomxy} in conservation form (using~\eqref{eq:ndcont}), which is given by
\begin{align}\label{eq:momxyda}
\bm{u}_t + \nabla_x\cdot\lrr{\bm{u}\otimes\bm{u}} + \lrr{w \bm{u}}_z &= -\nabla_x p  + f \bm{u}^* \nonumber \\
+ \Rey \mu \left( \nabla_x \cdot \begin{bmatrix} K_{11}&K_{12}\\K_{21}&K_{22} \end{bmatrix} \right. &  \left. + \partial_z\begin{bmatrix} \lrr{h_x K_{11} + h_y K_{12}}q \\ \lrr{h_x K_{21} + h_y K_{22}}q  \end{bmatrix} \right)   + O(\mu^3).
\end{align}
where $\otimes$ is the outer product, giving $(\bm{u} \otimes \bm{u})_{ij}=u_i u_j$. Averaging and simplifying this equation results in
\begin{align}\label{eq:momxyda2}
 & \lrr{h \overline{\bm{u}}}_t + \nabla_x \cdot \lrr{h \overline{\bm{u}\otimes\bm{u}}} -h \overline{\nabla_x p} - \nu\mu \lrr{\nabla_x \cdot \lrr{h \begin{bmatrix} \overline{K}_{11}&\overline{K}_{12}\\\overline{K}_{21}&\overline{K}_{22} \end{bmatrix}}} - fh\overline{\bm{u}^*} \nonumber\\
&= -\lrs{h_t \bm{u} + \nabla_xh \cdot \lrr{\bm{u}\otimes\bm{u} } + w\bm{u}}_{z=h} \nonumber \\
&+ \Rey \mu \lrs{-\nabla_x h \cdot \begin{bmatrix} K_{11}&K_{12}\\K_{21}&K_{22} \end{bmatrix}        + \begin{bmatrix} h_x K_{11} + h_y K_{12} \\ h_x K_{21} + h_y K_{22}   \end{bmatrix}}_{z=h}+ O(\mu^3),
\end{align}
where we have used that $h$ and $w$ are zero at $z=0$, and we have included the upper boundary terms in square brackets. Both of the boundary terms are in fact zero: the first follows from the kinematic boundary condition~\eqref{eq:fskbc}, while the second follows from the original construction of the stress tensor. 
Significantly, the odd-viscous contributions on the left-hand side in~(\ref{eq:momxyda2}) can be rewritten as a sum of two terms: 
\begin{equation}\label{eq:naked}
 \nu\mu h \nabla_x \cdot \begin{bmatrix} \overline{K}_{11}&\overline{K}_{12}\\\overline{K}_{21}&\overline{K}_{22} \end{bmatrix} + \nu\mu (\nabla_x h) \cdot  \begin{bmatrix} \overline{K}_{11}&\overline{K}_{12}\\\overline{K}_{21}&\overline{K}_{22} \end{bmatrix}.
\end{equation}
The first term in~(\ref{eq:naked}) has a typical viscous force due to a stress gradient in the bulk of the fluid. Surprisingly, the second term is not a stress gradient, and depends instead on the ``naked'' stress $\overline{K}$ as well as on the gradient of the surface profile $h$. This term highlights the effects of odd viscosity in the presence of a free surface, and cannot be observed in the fluid bulk.

Recalling the $z$-independence of the leading order terms for $\bm{u}$ in equation~\eqref{eq:expansion}, it can be shown that
\begin{align}
\nabla_x \cdot \lrr{h \overline{\bm{u}\otimes\bm{u}}} =  \nabla_x \cdot \lrr{h \overline{\bm{u}} \otimes \overline{\bm{u}}}  + O(\mu^3).
\end{align}
We wish to find the pressure gradient $\overline{\nabla_x p}$ in terms of $\overline{\bm{u}}$ and $h$. We do so by solving for the leading order $p\h{0}$ and $O(\mu^2)$ correction $p\h{2}$ to the pressure. Averaging the vertical momentum equation~\eqref{eq:ndmomz}, we find that at $O(1)$, combined with the dynamic boundary condition~\eqref{eq:ndfsdbc}, the leading order pressure is given by
\begin{align}\label{eq:p0}
p\h{0} &= h(x,y,t).
\end{align}
To evaluate $p\h{2}$, we must replace instances of $w$ in the vertical momentum equation~\eqref{eq:ndmomz} with terms of the form $\bm{\overline{u}}$, which is done by averaging the incompressibility condition~\eqref{eq:ndcont} to find $w = -z\lrr{\nabla_x \cdot \bm{\overline{u}}} + O(\mu^2)$. It follows that
\begin{equation}\label{eq:p2}
  p\h{2} = \frac{(z^2-h^2)}{2}  \lrs{\lrr{\partial_t +\lrr{\bm{\overline{u}} \cdot \nabla} } \nabla_x \cdot \bm{\overline{u}}  - \lrr{\nabla_x \cdot \bm{\overline{u}} }^2 } -  B \Delta_x h,
\end{equation}
Hence, using equations~\eqref{eq:p0} and~\eqref{eq:p2}, we finally recover the Serre nonlinearity
\begin{align}
 h \overline{\nabla_x p}
&=  h \nabla_x h - \mu^2 \nabla_x  \lrs{\frac{h^3}{3} \lrs{\lrr{\partial_t +\lrr{\overline{\bm{u}} \cdot \nabla} } \nabla_x \cdot \overline{\bm{u}} - \lrr{\nabla_x \cdot \overline{\bm{u}}}^2 } } + \mu^2 B h \nabla_x \Delta_x h  + O(\mu^3).
\end{align}
Substituting this into~\eqref{eq:momxyda2}, we recover the nonlinear system in conservation form for $\bm{u}$ and $h$, accurate up to $O(\mu^3)$, given by
\begin{align}
h_t + \nabla_x \cdot\lrr{h \overline{\bm{u}}} &=0, \label{eq:RSWE1} \\
\lrr{h \overline{\bm{u}}}_t + \nabla_x \cdot \lrr{h \overline{\bm{u}}\otimes\overline{\bm{u}}}
&=   fh\overline{\bm{u}^*}  - \frac12 \nabla_x h^2 + \Rey\mu \lrr{\nabla_x \cdot \lrr{h \begin{bmatrix} \overline{K}_{11}&\overline{K}_{12}\\\overline{K}_{21}&\overline{K}_{22} \end{bmatrix}}} \nonumber \\
+\mu^2 \nabla_x & \lrs{ \frac{h^3}{3} \lrs{\lrr{\partial_t +\lrr{\overline{\bm{u}} \cdot \nabla_x} } \nabla_x \cdot \overline{\bm{u}} - \lrr{\nabla_x \cdot \overline{\bm{u}}}^2 }} + \mu^2 B h \nabla_x \Delta_x h. \label{eq:RSWEcon2}
\end{align}
Equations \eqref{eq:RSWE1}-\eqref{eq:RSWEcon2} are three equations for three unknowns $(\eta,u,v)$. Equation \eqref{eq:RSWE1} is exact and enforces conservation of mass, while equation \eqref{eq:RSWEcon2} has errors of order $(\mu^3)$ and corresponds to conservation of momentum in which $z$-variations have been averaged.
Equation~\eqref{eq:RSWEcon2} can be written in convective form as
\begin{align}
\ol{\bm{u}}_t + \lrr{\ol{\bm{u}}\cdot\nabla_x}\ol{\bm{u}}  &=   f\overline{\bm{u}^*}  - \nabla_x h + \Rey\frac{\mu}{h} \lrr{\nabla_x \cdot \lrr{h \begin{bmatrix} \overline{K}_{11}&\overline{K}_{12}\\\overline{K}_{21}&\overline{K}_{22} \end{bmatrix}}} \nonumber \\
 + \frac{\mu^2}{h}  \nabla_x  &\lrs{\frac{h^3}{3} \lrs{\lrr{\partial_t +\lrr{\overline{\bm{u}} \cdot \nabla_x} } \nabla_x \cdot  \overline{\bm{u}} - \lrr{\nabla_x \cdot \overline{\bm{u}}}^2}   } + \mu^2 B  \nabla_x \Delta_x h. \label{eq:RSWE2}
\end{align}
The above system approximates the equations in Section~\ref{section:ref} with no assumption on the nonlinearity of the system. It is an odd-viscous extension of the Serre or Green-Naghdi equations which govern dissipation-free single-layer free-surface flows. 

\begin{figure}
	\centering
	\begin{overpic}[scale=1]{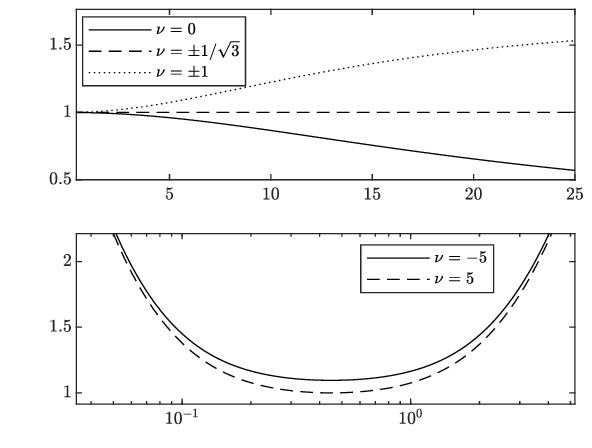}
		\put(0,72){$(a)$}
		\put(0,33){$(b)$}
		\put(53,0){$\lvert\bm{k}\rvert$}
		\put(53,40){$\lvert\bm{k}\rvert$}
		\put(3,24){$c_+$}
		\put(3,60){$c_+$}
	\end{overpic}
	\caption{Dispersion relation~\eqref{eq:disprel}   with $\mu=0.1$, $B=0$ and variable $\hat{f}$ and $\nu$. In panel $(a)$, we fix $\hat{f}=0$, and vary $\nu$, as shown in the legend. Note that the sign of $\nu$ does not effect the dispersion relation, and, curiously, for $\nu=\pm 1/\sqrt{3}$, the speed $c_+$ is constant. In panel $(b)$ of figure \ref{fig:disprel}, we fix $\hat{f}=1$, and vary $\nu$. From equation~\eqref{eq:disprel}, it can be seen that non-zero values of $\hat{f}$ introduce singularity at  $\lvert\bm{k}\rvert=0$. A change of sign in $\nu$ effects the dispersion relation, a consequence of changing the direction of internal rotations relative to external rotations. The plot in Figure~$(b)$ uses a logarithmic scale for the $\lvert\bm{k}\rvert$-axis.}
	\label{fig:disprel}
\end{figure}

\section{Linear and weakly nonlinear theory}\label{section:wnl}
In this Section, we will discuss the linear dispersion relation of this system, and derive weakly nonlinear theories to describe solitary waves one would expect for different parameter values.
\subsection{Linear theory}
Linearising the system, we  seek  wave-like solutions with wavenumbers $k_x$ and $k_y$ in the $x$ and $y$-direction. Denoting $\bm{k}=[k_x,k_y]^T$, we write
\begin{align}
\begin{bmatrix}
h-1 \\
u \\
v
\end{bmatrix}&=
\bm{A}
e^{i\lrr{\bm{k}\cdot \bm{x}-w t}} + \text{c.c.}, 
\end{align}
 where c.c. stands for complex conjugate. 
Solving the linear system, one recovers the isotropic dispersion relation, which has three branches $\omega_+$, $\omega_-$ and $\omega_0$, given by
\begin{align}\label{eq:disprel}
%w_{\blue{\pm}} &= \blue{\pm} \lrs{ \frac{\lvert \bm{k} \rvert^2 +\mu^2\lrr{\lrr{\Rey^2\blue{+B}}\lvert\bm{k}\rvert^4 - 2\Rey \hat{f} \lvert \bm{k} \rvert^2  + \hat{f}^2}}{1+\frac{1}{3}\mu^2 \lvert \bm{k}\rvert^2}}^{1/2}, && \text{or} \,\,\,\,\,\, w_{\blue{0}}=0,
\omega_{\pm} &= \pm \lrs{ \frac{\lvert \bm{k} \rvert^2 +\mu^2 B \lvert\bm{k}\rvert^4 + \mu^2 \lrr{ \Rey \lvert \bm{k} \rvert^2  -\hat{f}}^2}{1+\frac{1}{3}\mu^2 \lvert \bm{k}\rvert^2}}^{1/2}, && \text{or} \,\,\,\,\,\, \omega_{0}=0,
\end{align}
%where the phase velocity is $c_{\blue{\pm}}=w_{\blue{\pm}}/|\bm{k}|$
where the rescaled Coriolis parameter $\hat{f}$ is given by
\begin{equation}
    \hat{f} = \frac{f}{\mu}.
\end{equation}
The branch $\omega_{0}=0$ are the so-called inertial waves, with solutions given by 
\begin{align}
\begin{bmatrix} 
h-1 \\
u \\
v
\end{bmatrix}&=
\begin{bmatrix} 
\mu\lambda/\alpha_B\\ ik_y \\ -i k_x
\end{bmatrix} A e^{i\bm{k}\cdot \bm{x}}+ \text{c.c.} && \text{where} \,\,\, \lambda = \nu \lvert \bm{k} \rvert^2 - f, \hspace{.5cm} \alpha_B=1+\mu^2 B \lvert \bm{k} \rvert^2,
\end{align} 
where $A$ is an arbitrary constant. Inertial waves requires rotation, either external (Coriolis effect) or internal (odd-viscous), to exist. 

Next, consider the branches $\omega_\pm$. When the Coriolis force is ignored ($\hat{f}=0$), the effects of odd viscosity are qualitatively similar to that of surface tension. An odd-viscous fluid (without surface tension) would result in the same dispersion relation as a classical fluid (without odd viscosity) provided that the Bond number is given by $B=\nu^2$. This is not true for non-zero $\hat{f}$  due to the coupling of the Coriolis force and odd viscosity, highlighting the chiral nature of these terms. Because the equations are rotationally symmetric with an isoptropic dispersion relation, without loss of generality, we consider a wave travelling in the $x$-direction $(k_y=0)$. The corresponding solution is
 \begin{align}\label{eq:lin3}
\begin{bmatrix} 
h-1 \\
u \\
v
\end{bmatrix}&=
\begin{bmatrix} 
1 \\c_\pm   \\ \frac{i\mu \lambda}{k}%\frac{i}{k_x\mu\lambda } \lrr{w_{\pm}^2 ( 1+ \frac{1}{3}\mu^2k_x^2)- k_x^2 \alpha_B }
\end{bmatrix} A e^{ik_x (x- c_\pm t)}+ \text{c.c.}
\end{align}
where $A$ is an arbitrary constant and $c_\pm=\omega_\pm/k$ is the phase speed. Note that while the dispersion relation is clearly isotropic, the chirality is reflected in the linear modes. In particular, time-reversal symmetry requires a change of sign for $\nu$ and $f$, i.e., the direction of internal and external rotations. In other words, the linear modes are invariant under $t \rightarrow -t$, $c \rightarrow -c$, $(u,v) \rightarrow -(u,v)$, $\nu\rightarrow -\nu$ and $\hat{f}\rightarrow-\hat{f}$. 
 %(i.e. to reverse the direction of internal and external rotations),  and hence the flow perpendicular to wave propagation $v$ reverses direction. Hence, although the dispersion relation $\omega_\pm$ is invariant under this mapping, the corresponding eigenvalue is not. 
The dispersion relation $\omega_\pm$ is invariant under a reversal of the sign of internal and external rotations alone, but the flow is not. A solution with no internal or external rotation ($\lambda=0$) exhibits flows in the direction of wave propagation only.

When the model is considered in the short-wavelength limit, the waves are non-dispersive to leading order, for arbitrary parameters. To see this, one may take the limit $\lvert\bm{k}\rvert \rightarrow \infty$ above or follow the derivation of so-called Avron waves~\cite{avron1998odd} in fluids with odd viscosity, but including the dispersive correction to hydrostatic pressure. Coincidentally, the dispersion of Avron waves exactly cancels the leading-order gravitational dispersion at short wavelengths. For plane waves travelling uni-directionally in the $x$-direction, this results in a non-dispersive wave equation,
\begin{align}
h_{tt} = 3(\nu^2+B) h_{xx},
\end{align}
with wave speed $c = \sqrt{3(\nu^2+B)}$.
The absence of dispersion at short wavelengths means less energy can disperse during nonlinear steeping, increasing the likely-hood of shock solutions \cite{zeitlin2007nonlinear,lahaye2012shock}.

%enhances the possibility of shock solutions \cite{lahaye2012shock}.} 

%\red{REFEREE 1: WHAT DOES ENHANCE MEAN}

% We restrict our attention to $w_+$, noting that $w_-=-w_+$.} 

Figure~\ref{fig:disprel} shows the dispersion relation $c_+$ for different values of the parameters, fixing both $\mu=0.1$ and $B=0$ (i.e. no surface tension). We plot only the positive root of the dispersion relation. In panel $(a)$, we remove the Coriolis effect by setting $\hat{f}=0$, and vary $\nu$. The sign of $\nu$ does not affect the dispersion relation, as only the square of $\nu$ appears in equation \eqref{eq:disprel} when there are no external rotations ($\hat{f}=0$). The odd-viscous term does not affect the phase velocity for long wavelengths (that is, $\lim_{k\rightarrow 0} c_{+}$). There is a critical value of $\nu=\nu^*=\pm 1/\sqrt{3}$ at which  the dispersion relation changes from monotonically decreasing for $\nu<\nu^*$ to monotonically increasing for $\nu>\nu^*$. The case $\nu=\nu^*$ results in the curious situation that $c_{+}=1$, and the system is no longer dispersive to the order considered. In this case, the dispersive effects of odd viscosity balance the finite-depth corrections to the dispersion at $O(\mu^2)$, a case similar to shallow-water gravity-capillary waves when the Bond number is $1/3$~\cite{milewski1999time}.
 In panel $(b)$, we fix $\hat{f}=1$ and again vary $\nu$. The effects of Coriolis forces dominate long wavelengths, such that equation~\eqref{eq:disprel} is singular as $\lvert\bm{k}\rvert \rightarrow  0$ with the scaling $ c_{+} \sim {\mu \hat{f}}{\lvert\bm{k}\rvert^{-1}}$. The sign of $\nu$ effects the dispersion relation for non-zero $\hat{f}$, since the direction of internal rotations relative to external rotations changes.

\begin{figure}
    \centering
    %\begin{overpic}[scale=1]{figs_p/min_reg.pdf}
    \begin{overpic}[width=11cm]{figs_p/test2.jpg}
    \put(4,19){$\hat{f}$}
    \put(52,19){$\hat{f}$}
    \put(0,31){$(a)$}
    \put(50,31){$(b)$}
    \put(28.5,0){$\nu$}
    \put(77,0){$\nu$}
    \put(26,35){$B=0$}
    \put(74,35){$B=1$}
    \end{overpic}
    \caption{The figure shows, in grey, the regions of parameter space in which a minimum of $c_+$ occurs. Both panels have $\mu=0.2$, while panel $(a)$ and $(b)$ have $B=0$ and $B=1$ respectively. Note that $\hat{f}\rightarrow 0$ is a singular limit, and there is never a minimum for $\hat{f}=0$.  }
    \label{fig:min_reg}
\end{figure}

Solitary waves bifurcate from points where the phase and group velocities are equal.
Furthermore, except for the special case of embedded solitary waves \cite{champneys2001embedded}, they are typically found outside the linear spectrum. When $\hat{f}=0$, the group and phase velocity are equal at $\lvert \bm{k} \rvert =0$ (i.e., the long-wave speed), where $c_{+}=1$. Hence, one may expect to find \emph{long-wave} solitary waves bifurcating from zero amplitude at $c_{+}=1$, and the speeds of the waves will be greater than unity for $\nu<\nu^*$, and less than unity for $\nu>\nu^*$.
On the other hand, when there are also external rotations, the singular behaviour of $c_{+}$ at $\lvert \bm{k} \rvert=0$ removes the possibility of finding a solitary wave bifurcation about this point. However, there is another candidate for solitary wave bifurcations, at a global minimum of the dispersion relation at $|k|=k_m$, denoted $c_m$. These are called \emph{wavepacket} solitary waves. It can be checked by direct calculation that the group and phase velocities are equal at this point \cite{akylas1993envelope}, and furthermore for speeds $-c_m<c<c_m$ there are no linear waves (i.e., there is a gap in the linear spectrum for $-c_m<c<c_m$). Seeking solutions to $dc_{+}/dk=0$, we find that $k_m$ satisfies
\begin{align}
\lrr{1-3\lrr{\nu^2+B}-2\mu^2\nu \hat{f}}k_m^4 + 2\mu^2 \hat{f}^2 k_m^2 + 3\hat{f}^2 =0.
\end{align}
For a minimum to exist, we require real solutions for $k_m$, which occurs under the condition 
\begin{equation}\label{eq:minex}
3\lrr{\nu^2+B}-1 + 2 \mu^2 \nu \hat{f} >0, \qquad \hat{f}\neq 0
\end{equation}
The above condition can be written as
\begin{align}
\begin{cases}
\hat{f} > \frac{1-3(\nu^2+B)}{2 \mu^2 \nu}, \,\,\,\, \hat{f}\neq 0,  & \text{when} \,\,\,\,\, \nu>0  \\
\hat{f} < \frac{1-3(\nu^2+B)}{2 \mu^2 \nu}, \,\,\,\, \hat{f}\neq 0, & \text{when} \,\,\,\,\, \nu<0 \\
B>\frac{1}{3} \,\,\,\, \hat{f}\neq 0 & \text{when} \,\,\,\,\, \nu=0, 
\end{cases}    
\end{align}
Fixing $\mu=0.2$, figure \ref{fig:min_reg} shows the parameter regions for which a minimum of the dispersion exists for $B=0$ and $B=1$ for panel $(a)$ and $(b)$ respectively. For $B<1/3$, a minimum occurs does not occur for any value of $\hat{f}$ given $\nu=0$, as demonstrated in panel $(a)$. Hence, for odd-viscous waves without surface tension, internal rotations are required for a minimum. For Bond numbers greater than $1/3$, a minimum occurs with $\nu=0$ given $\hat{f}\neq0$ ($\hat{f}\rightarrow 0$ is a singular limit). For non-zero $\nu$, the dispersion relation typically has a minimum given the external rotations are of the same sign as $\nu$. That is, internal and external rotations in the same direction create preferential conditions for a dispersion relation minimum, and hence the possibility of localised wavepacket solutions. %This information may be of use for practitioners who wish to explore localised wave behaviour and interactions in externally rotating three-dimensional chiral fluids, analogous to those explored in discrete mediums} \red{(CITE sol-wave papers in fibre optics).  }

The bifurcation of wavepacket solitary waves from zero amplitude requires the additional condition that the corresponding nonlinear Schr\"{o}dinger equation for modulations of monochromatic waves at $k_m$ is of focusing type. The bifurcation structure described above can be predicted by weakly nonlinear theories, which we present in the following section.

\subsection{Weakly nonlinear theory}
Weakly nonlinear, weakly dispersive systems can be recovered by suitable scalings. We seek unidirectional models, and without loss of generality choose the waves to travel in the positive $x$-direction. To consider weakly nonlinear theory, we rescale the system as follows
\begin{align}
\bm{u} &= \epsilon \hat{\bm{u}}   && h = 1 + \epsilon \eta
\end{align}
with $\epsilon \ll 1$.
We consider the classical Boussinesq scaling relating the shallow-water parameter to the amplitude, given by $\mu^2=\epsilon$. Furthermore, we take a frame of referencing moving with the long wave speed via a Galilean transform, given by $X=x-(1-\hat{f}\nu\mu^2)t$, and consider a slowly varying time variable $T=\epsilon t$.
\begin{figure}
    \centering
    \begin{overpic}[scale=1]{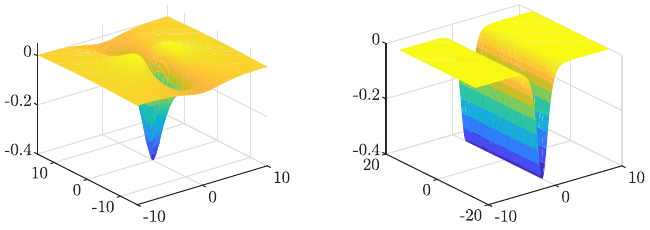}
    \put(-2,29){$(a)$}
    \put(-4,18){$\eta_0$}
    \put(5,2){$Y$}
    \put(33,1){$X$}
    \put(51,29){$(b)$}
   \put(50,18){$\eta_0$}
    \put(60,2){$Y$}
    \put(88,1){$X$}
    
    \end{overpic}
    \caption{Panels $(a)$ and $(b)$ show the solitary wave solutions~\eqref{eq:KP_sol} and~\eqref{eq:KDV_sol} with $A=-0.4$, $\nu^2+B=4/9$, and $\hat{f}=0$. }
    \label{fig:KP_KDV}
\end{figure}
To consider slow variance in the $y$-direction, we introduce a new $y$-scaling, given by
\begin{align}
Y=\sqrt{\epsilon} y.
\end{align}
We seek the prefactors in a power series of $\sqrt{\epsilon}$, that is
\begin{align}
\eta &= \eta_0 + \sqrt{\epsilon} \eta_1 + \epsilon \eta_2 + \cdots, \\
\bm{\hat{u}} &= \bm{\hat{u}_0} +  \sqrt{\epsilon} \bm{\hat{u}_1} + \epsilon \bm{\hat{u}_2} + \cdots,
\end{align}
where $\bm{\hat{u}}=[\hat{u},\hat{v}]$.
At leading order, one recovers the linear dispersion relation. At $O\lrr{\sqrt{\epsilon}}$, one recovers the solution $\hat{v}_1$. It is found at $O\lrr{\sqrt{\epsilon}}$ that the functions  $\hat{\eta}_1$ and $\hat{u}_1$ are arbitrary and can be absorbed into the definition of  $\eta_0$ and  $\hat{u}_0$. At  $O\lrr{\epsilon}$, one recovers a solvability condition for $\eta_0$. %The leading order solutions are then given by
The solutions up to $O(\sqrt{\epsilon})$ are then given by
\begin{align}\label{eq:los}
\eta_0&=\hat{u}_0, && \hat{v}_0=0, && \eta_1=\hat{u_1}=0 && \hat{v}_{1X}= \eta_{0Y} + \nu \eta_{0XX} + \hat{f}\eta_0,
\end{align}
where the function $\eta_0$ satisfies the equation
\begin{align}
\lrs{2\eta_{0T}  + 3 \eta_{0}\eta_{0X} + \lrr{\frac{1}{3}-\nu^2  -B}\eta_{0XXX}}_X  &= \hat{f}^2 \eta_{0} - \eta_{0YY} \label{eq:RMKP}. 
\end{align}
Equation~\eqref{eq:RMKP} is the rotation-modified Kadomtsev-Petviashvilli (KP) equation or Melville-Grimshaw equation~\cite{grimshaw1989derivation}. The solvability condition was recovered at $O(\epsilon)$, resulting in the approximation being valid with errors of order $O(\epsilon^{3/2})$. The odd viscosity has the effect of modifying the coefficient of the dispersive term in the direction of travel.
Seeking linear perturbations of the form $\eta_0\sim e^{i(k_x X+k_y Y - cT)}$, the dispersion relation $c(k_x,k_y)$ for equation~\eqref{eq:RMKP}  is singular at $k_x=k_y=0$, in agreement with the full system~\eqref{eq:disprel}. Therefore, one does not expect to find solitary waves bifurcating from zero amplitude about $|\bm{k}|=0$.

Removing the Coriolis effect, we recover the KP equation
\begin{align}
\lrs{2\eta_{0T} + 3 \eta_{0}\eta_{0X} + \lrr{\frac{1}{3}-\nu^2 -B}\eta_{0XXX}}_X  &=  -  \eta_{0YY} \label{eq:KP}.
\end{align}
Depending on the sign of the dispersive terms, it is known as the KP1 equation (for $\nu^2+B>1/3$) or the KP2 equation (for $\nu^2+B<1/3$).
The KP1 equation has travelling wave solutions which are localised in both dimensions, and which are known as lump solitons. These solutions bifurcate from $\lvert \bm{k} \rvert=0$ \cite{manakov1977two}. Denoting the speed of propagation by $c$, lump soliton solutions are given by
\begin{align}\label{eq:KP_sol}
\eta_0 &= A \left[ \frac{\frac{3A}{8(3\nu^2+B -1)} X^2 + \frac{9A^2}{64(3\nu^2 +B -1)} Y^2 + 1 }{\left(-\frac{3A}{8(3\nu^2 +B -1)} X^2 + \frac{9A^2}{64(3\nu^2 +B -1)} Y^2 + 1\right)^2 }\right],  && c=1+\frac{3}{16}A,
\end{align}
where $A<0$ is a free constant. 
A KP1 soliton with $\nu^2+B=4/9$ is shown in panel $(a)$ of Figure~\ref{fig:KP_KDV}.

We now consider plane waves, that is solutions with invariance in the $y$-direction. We denote $k \equiv k_x$ as the wavenumber along the direction of propagation.
The governing equation is the Ostrovsky equation, given by
\begin{align}
\lrs{2\eta_{0T} + 3 \eta_{0}\eta_{0X} + \lrr{\frac{1}{3}-\nu^2 -B} \eta_{0XXX}}_X  &= \hat{f}^2 \eta_{0} \label{eq:OS}. 
\end{align}
Like the rotation-modified KP equation, the Ostrovsky equation does not admit soliton solutions about $k=0$, due to the singular nature of $c$ there.  The work of Ref.~\cite{grimshaw2016formation,obregon1998oblique} found solitary wave solutions bifurcating about the minimum of the dispersion relation when the dispersive term is negative. The linear dispersion relation for the Ostrovsky equation in the original coordinate system, which we denote $c^{\text{o}}$, is given by
\begin{equation}\label{eq:disprelostr}
c^{\text{o}} = 1 - \hat{f}\nu\mu^2 + \mu^2\left(\frac{\hat{f}^2}{2k^2} - \frac{1}{2}\lrr{\frac{1}{3}-\nu^2 -B} k^2 \right).
\end{equation}
with a minimum at $k=k_m^{\text{o}}$, where
\begin{equation}
k_m^{\text{o}} = \left(\frac{\hat{f}^2}{(\nu^2+B)-\frac{1}{3}}\right)^{1/4}.
\end{equation}
Therefore, the existence of a minimum requires $\nu^2+B>1/3$, in agreement with the condition for the Serre system~\eqref{eq:minex} when $\mu=0$. In fact, expanding~\eqref{eq:disprel} in powers of $\mu$, we find that the linear dispersion relation is equivalent to that of the Ostrovsky equation~\eqref{eq:disprelostr} at $O(\mu^2)$. Therefore, given $\mu\ll 1$, one may expect good agreement between the models for weakly nonlinear solutions. This is explored further in section \ref{section:fneq0}.

Whereas the KdV equation (see below) and KP equations are appropriate to describe the solitary wave bifurcation at zero wavenumber, the behaviour about a finite wavenumber $k$, for plane waves, is described by a one-dimensional nonlinear Schr\"{o}dinger (NLS) equation. One arrives at the NLS equation by seeking a slowly modulated wavepacket with carrier wave of wavenumber $k$ and wavepacket amplitude $A$. The multiscale modulation theory here is valid as it describes a dynamics where the carrier wave is long relative to the depth of the fluid and its modulation is long relative to the carrier wave. For simplicity, we do not consider the effects of surface tension ($B=0$). Denoting $\epsilon$ to be a small parameter, and given the packet varies slowly in time (depending on $\tau=\epsilon^2 t$) and travels with group velocity $c_g$ (depending on $\xi=\epsilon(x-c_g t)$), the governing equation for the wavepacket amplitude $\epsilon A(\xi,\tau)$ is given by 
\begin{align}\label{eq:NLS}
iA_\tau + \alpha A_{\xi\xi} = \beta \lvert A \rvert^2 A,
\end{align}
where $\alpha$ and $\beta$ are given by equations~\eqref{eq:NLS_ap}--\eqref{eq:beta}.
A derivation of the above equation is found is Appendix~\ref{section:NLS}.
It is known that the NLS admits `bright' solitary wavepackets when of the focusing type (that is, $\alpha\beta  <0$) and `dark' solitary waves with oscillatory tails when of the defocussing type (that is, $\alpha\beta >0$), when the linear group and phase velocity are equal at the chosen wavenumber $k$. The algebra quickly becomes unwieldy when trying to check whether the NLS equation is focussing or defocussing at the dispersion relation minimum (where $c_g=c$). However, one can express $\alpha$ and $\beta$ to leading order in $\mu$ for $\mu\ll 1$ to find 
\begin{align}\label{eq:alphabeta}
%k_0^2 = \frac{\lvert f \rvert}{\sqrt{\nu^2-\frac{1}{3}}}, &&
\alpha_0 &= 2 \mu^2 \lvert f \rvert^{1/2} \left(\nu^2-\frac{1}{3}\right)^{3/4}, && \beta_0  = - \frac{1}{\mu^2} \left(\nu^2-\frac{1}{3}\right)^{-1},   
\end{align}
 where $\alpha_0$ and $\beta_0$ are the leading order values of $\alpha$ and $\beta$ at the dispersion relation minimum ($k_m$). The condition for a minimum to exist is given by equation \eqref{eq:minex} and, at leading order in $\mu$, gives $\nu^2>1/3$. Hence, from equations \eqref{eq:alphabeta} we find that $\alpha>0$ and $\beta<0$ to leading order, and the corresponding NLS is of the focusing type at the minimum. In addition, these expressions provide the scalings for nonlinearity and dispersion to balance in the shallow water limit, with $A \sim \mu^2$ providing that balance.
 
 %Furthermore,}
%\blue{numerically sweeping parameters, for all $\nu$, $\hat{f}$ in the range $[-10,10]$ and $\mu$ in $(0,10]$, if a minimum in the dispersion relation exists, it was found that the corresponding equation~\eqref{eq:NLS} is of the focusing type.}

\begin{figure}
    \centering
    \begin{tikzpicture}
	\node[anchor=south west,inner sep=0] (image) at (0,0) {\includegraphics[scale=1]{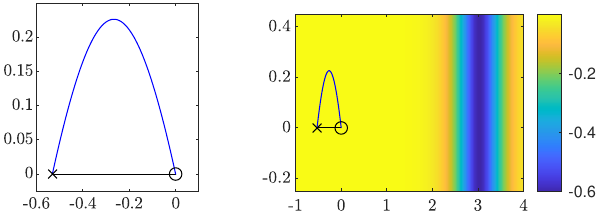}};
	\begin{scope}[x={(image.south east)},y={(image.north west)}]
	\draw [ -{Triangle[open]}]  (0.6,0.97) -- node [midway,above] {Direction of propagation} (0.8,0.97);
		\draw [ solid, ->, thick]  (0.625,0.2) -- node [above,yshift=0.2cm] {$y$} (0.625,0.325);
	\draw [ solid, ->, thick]  (0.625,0.2) --  node [right,xshift=0.2cm] {$x$} (0.675,0.2);
	\node [] () at (0.18,-0.02) {$x$};
	\node [] () at (0.0,0.53) {$y$};
	\end{scope}
	\end{tikzpicture}
    \caption{Particle path for a KdV solitary wave with odd viscosity ($\nu=2/3$, blue curve) and surface tension ($B=4/9$, black curves). The solution has parameters $A=-2$ and $\epsilon=0.3$. Panel $(a)$ shows the particle path after a solitary wave has passed through, where the start and the end of the path are given by the circle and the cross, respectively. Panel $(b)$ shows the wave in the $xy$-plane, where the colour bar corresponds to the value of $u$.}
    \label{fig:KDV_PP}
\end{figure}
\begin{figure}
    \centering
  	\begin{tikzpicture}
	\node[anchor=south west,inner sep=0] (image) at (0,0) {\includegraphics[scale=1]{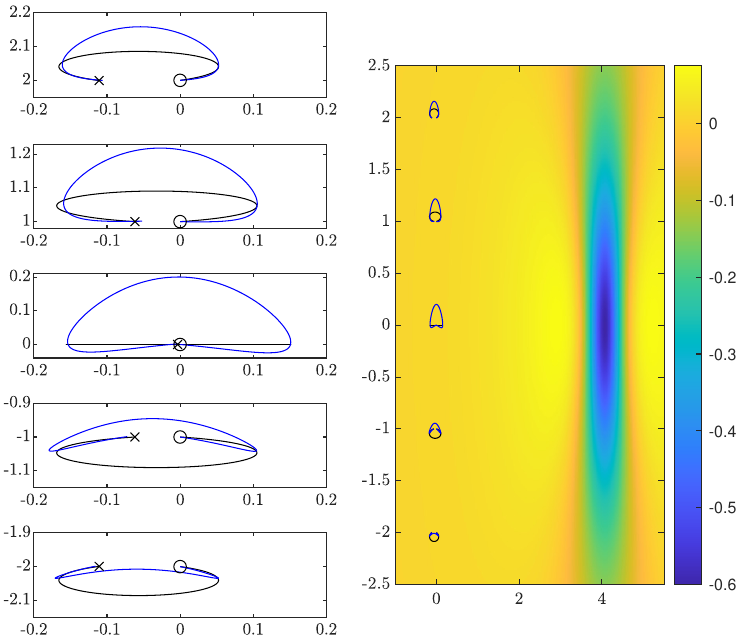}};
	\begin{scope}[x={(image.south east)},y={(image.north west)}]
	\draw [ solid, ->]  (0.475,0.88) --  (0.56,0.84);
	\draw [ solid, ->]  (0.475,0.7) --  (0.56,0.68);
	\draw [ solid, ->]  (0.475,0.55) --  (0.56,0.52);
	\draw [ solid, ->]  (0.475,0.275) --  (0.56,0.32);
	\draw [ solid, ->]  (0.475,0.125) --  (0.56,0.16);
	\draw [ -{Triangle[open]}]  (0.7,0.92) --  node [midway,above] {Direction of propagation}  (0.8,0.92);
	\draw [ solid, ->, thick]  (0.7,0.15) -- node [above,yshift=0.3cm] {$y$} (0.7,0.2);
	\draw [ solid, ->, thick]  (0.7,0.15) --  node [right,xshift=0.3cm] {$x$} (0.75,0.15);
	\node [] () at (0.24,-0.02){$x$};
	\node [] () at (-0.01,0.1) {$y$};
	\node [] () at (-0.01,0.3) {$y$};
	\node [] () at (-0.01,0.5) {$y$};
	\node [] () at (-0.01,0.7) {$y$};
	\node [] () at (-0.01,0.9) {$y$};
	\end{scope}
	\end{tikzpicture}
	  \caption{Particle paths for a KP solitary wave~\eqref{eq:KP_sol} with odd viscosity ($\nu=2/3$, blue curves) and surface tension ($B=4/9$, black curves). The solution has parameters $A=-2$ and $\epsilon=0.3$.  Five particle paths are shown in the left-hand panels. The particles start at the circle, and end at the cross. Their position in relation to the solitary wave is shown in the right-hand panel, where the colour bar corresponds to the value of $u$.        \label{fig:KP_PP}}
\end{figure}

Finally, considering plane waves without the Coriolis effect, we recover the celebrated Korteweg-de Vries (KdV) equation. In the case of vertical odd viscosity that we consider, this equation has the form
\begin{align}
2\eta_{0T} + 3 \eta_{0}\eta_{0X} + \lrr{\frac{1}{3}-\nu^2-B}\eta_{0XXX}  &= 0 \label{eq:KdV}. 
\end{align}
The KdV equation admits the famous $\sech^2$ soliton solutions about $k=0$ for all parameters where the dispersive coefficient is non-zero. These solutions are given explicitly by 
\begin{align}\label{eq:KDV_sol}
\eta_0 &= A \sech^2 \lrs{ \lrr{\frac{A}{12(1-3(\nu^2+B))}}^{1/2} X  },&& c=1+\frac{A}{2}\mu^2.
\end{align}
When $\nu^2+B<1/3$, the solitons are waves of depression ($A<0)$, while for $\nu^2+B>1/3$ the solutions are waves of elevation ($A>0$).
One such soliton is shown in Figure~\ref{fig:KP_KDV}$(b)$.

Monteiro \textit{et. al.}~\cite{monteiro2020non} recently derived the KdV equation for two-dimensional surface water waves with odd viscosity. They find a dispersive coefficient due to odd viscosity which is linear in $\nu$, instead of the quadratic prefactor $\nu^2$ in equation~\eqref{eq:KdV}. Furthermore, they found the contribution to the dispersive coefficient due to odd viscosity has different signs for right-moving and left-moving waves. A consequence of this is that right and left movers have different free-surface perturbations for the same value of $\nu$. This is strikingly different behaviour to the model we consider, and stems from the different choices of the axis for the internal rotations. In the geometry considered by Monteiro \textit{et. al.}, odd viscosity is induced via vortex filaments with axes of rotation which are perpendicular to both the direction of propagation and the free-surface normal. This introduces a handedness, creating a preferred coordinate system. In the three-dimensional geometry we consider, vortex filaments have an axis of rotation which points along the free-surface normal (appendix \ref{appendix:OV}). Hence, in our case, there is no preferred coordinate system related to free-surface perturbations. However, in our case, the velocity vector is perpendicular to the axis of rotation, and consequently the chiral nature of the rotations affects the flow velocity. %\red{1. We don't change sign for left vs right-travelling and 2. time-reversability?}     \red{... This ..... Still, odd viscosity of this type has lead to interesting phenomena, such as breaking bulk boundary correspondence (cite) and  ....}

When $\hat{f}=0$, the effect of odd viscosity appears similar to that of a surface tension, but interesting differences are found within the structure of the flow. In particular, the leading order velocity perpendicular to the direction of wave propagation, $\hat{v}_1$ (given by equation~\eqref{eq:los}), has a contribution due to the odd viscosity. This is demonstrated in figure~\ref{fig:KDV_PP}, where we plot a particle path for an $x$-dependent KdV soliton~\eqref{eq:KDV_sol} with $\nu=2/3$ and $B=0$. The particle path is shown in blue, while the black curve corresponds to a particle path for a gravity-capillary KdV soliton with $\nu=0$ and $B=4/9$. We note that the interfaces are identical and the particle trajectories end in the same position. However, unlike the gravity-capillary wave, the flow arising from the KdV solution with odd viscosity has non-zero velocity perpendicular to the direction of wave propagation.

%We note here that the weakly nonlinear equations discussed above can be derived for gravity-capillary surface waves, where as mentioned earlier the term $\nu^2$ is replaced by the Bond number $B$. Hence, the profile of the leading order solution $\eta_0$ for the weakly nonlinear theory is the same for gravity-capillary waves and  odd-viscous waves with $\nu=B^{1/2}$. However, an interesting difference is found in the structure of the flow. In particular, the leading order velocity perpendicular to the direction of wave propagation is $\hat{v}_1$ (given by equation~\eqref{eq:los}) has a contribution due to the odd viscosity. 

In Figure~\ref{fig:KP_PP}, we plot particle paths for an $xy$-dependent KP1 soliton~\eqref{eq:KP_sol} with $(\nu,B)=(2/3,0)$ in blue and $(\nu,B)=(0,4/9)$ in black. While for gravity-capillary waves, the trajectories are reflected about $y=0$, this symmetry is violated for the odd-viscous waves. In particular,  particle paths for the odd-viscous wave above $y=0$ have a stronger perpendicular velocity in the positive $y$-direction, while the perpendicular velocities due to the KP $\eta_{0Y}$ term in equation~\eqref{eq:los} compete with those of odd viscosity for $y<0$. The odd viscosity enters the weakly nonlinear equations in the form $\nu^2$ (see equation~\eqref{eq:RMKP}), and hence changing the sign of $\nu$ does not affect the profile of the solitary wave. However, it does change the sign of the contribution to the perpendicular velocity $\hat{v}_1$ in equation~\eqref{eq:los}. Hence, the weakly nonlinear system retains symmetry under overall time reversal when signs of both the direction of propagation and the odd viscosity are flipped, that is under $c \to - c$ and $\nu \to - \nu$. 

\section{Nonlinear computations}
In the previous section, weakly nonlinear reductions of the Serre system (\ref{eq:RSWE1}--\ref{eq:RSWE2}) were discussed. As a check on the range of validity of these approximations, in this section we compute travelling solitary wave solutions to the odd-viscous Serre equations, and compare the results with those of Section~\ref{section:wnl}. We restrict our attention to one-dimensional plane waves, and the effects of surface tension are ignored.
\begin{figure}
    \centering
    \begin{overpic}[]{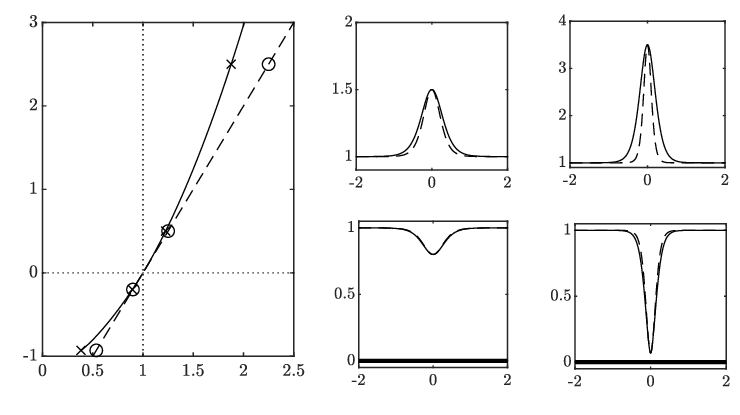}
    \put(0,52){$(a)$}
    \put(41,52){$(b)$}
    \put(70,52){$(c)$}
    \put(41,24){$(d)$}
    \put(70,24){$(e)$}
    \put(24,23){$(b)$}
    \put(26,48){$(c)$}
    \put(12.5,17){$(d)$}
    \put(7,11){$(e)$}
    \put(0,28){$A$}
    \put(23,2){$c$}
    \put(58,28){$x$}
    \put(87,28){$x$}
    \put(58,1){$x$}
    \put(87,1){$x$}
    \put(42,43){$h$}
    \put(71,43){$h$}
    \put(42,17){$h$}
    \put(71,17){$h$}
    \end{overpic}
    \caption{Panel $(a)$ shows two solitary wave branches with $\mu=0.1$ and $f=0$. The parameter $A$ is given by equation~\eqref{eq:A}. The depression ($A<0$) and elevation ($A>0$) branches  have  $\Rey=0.25$ and $\Rey=1$, respectively. The solid curves are solutions to the Serre equations, while the dashed curves are steady KdV travelling waves~\eqref{eq:KDV_sol}. The dotted curves are given by $A=0$ and $c=1$. Both branches bifurcate from zero amplitude at $c=1$. The solid and dashed lines in panels $(b)$--$(e)$ correspond to the Serre and KdV solutions highlighted with a cross and circle respectively in  panel $(a)$. }
    \label{fig:FNsoln}
\end{figure}
We impose invariance in the $y$-direction and find that the system reduces to
\begin{align}
h_t + \lrr{u h}_x &=0,  \label{eq:planewave1}  \\
u_t + u u_x  &= -h_x + \Rey\mu\lrr{\frac{h_xv_x}{h} + v_{xx}} +  \frac{\mu^2}{3h} \lrs{h^3 \lrr{u_{xt} + u u_{xx} - u_x^2 }}_x - \mu \hat{f} v, \label{eq:planewave2} \\
v_t + u v_x  &=  -\Rey\mu\lrr{\frac{h_xu_x}{h} + u_{xx}} +  \mu \hat{f} u. \label{eq:planewave3} 
\end{align}
We seek solutions travelling to the right with constant speed $c$ ($\partial_t \rightarrow -c\partial_x$) which decay at infinity, that is
\begin{align}\label{eq:decay}
h &\rightarrow 1, && u \rightarrow 0, &&  v \rightarrow 0,  && \text{as} \hspace{0.2cm} x\rightarrow \pm\infty.
\end{align}
We integrate the conservation of mass equation~\eqref{eq:planewave1} to find the constraint
\begin{align}\label{eq:planewave11}
h = \frac{c}{c-u}.
\end{align}
Furthermore, the momentum equation in the direction perpendicular to wave propagation~\eqref{eq:planewave3} becomes
\begin{equation}\label{eq:planewave33}
v_x = \frac{\Rey\mu}{c-u} \lrr{\frac{h_x u_x}{h} + u_{xx}} + \mu \hat{f} \lrr{h - 1}.
\end{equation}
Integrating~\eqref{eq:planewave33}, and using~\eqref{eq:planewave11}, we find that
\begin{align}\label{eq:vint}
v(x)= \mu \Rey \frac{h_x}{h} + \mu \hat{f} \int_{-\infty}^x (h-1) \, \mathrm{d}x.
\end{align}
Having solved the conservation of mass and the $y$-momentum equation to recover $h$ and $v$ explicitly in terms of $u$ and its derivative, we proceed to numerically solve the $x$-momentum equation~\eqref{eq:planewave2}. The code is written in MATLAB. The integral in the expression for $v$~\eqref{eq:vint} is numerically approximated using the trapezoidal rule. 
We take a periodic domain with length $L$, and use a pseudospectral collocation method  with $N$ equally spaced meshpoints, utilizing MATLAB's fast Fourier transform (FFT) routine. A typical number of meshpoints is $N=2^{11}$. The system is solved using the Newton-Raphson method, and we say a solution is converged once the $L^\infty$-norm of the residuals is of the order $\sim10^{-11}$. We choose a domain size sufficiently large such that the solutions become invariant to further increase in the domain size. This is checked by computing a solution with double the domain size and checking that the $L^\infty$-norm of the difference between the two solutions is of the order same order as the tolerance the residuals, that is $\sim10^{-11}$.

\subsection{Solitary waves for $f=0$}
When $f=0$, the ordinary differential equation for $u$ reduces to 
\begin{align}
-cu_x + u u_x  &= -\frac{u_x}{(c-u)^2} + \Rey^2\mu^2\lrr{\frac{u_x}{c-u} + \frac{\partial}{\partial x}} \lrr{\frac{u_x^2}{(c-u)^2} + \frac{u_{xx}}{c-u}} \nonumber \\
&+  \frac{\mu^2}{3h} \lrs{h^3 \lrr{-cu_{xx} + u u_{xx} - u_x^2 }}_x . \label{eq:OVnew}
\end{align}
We contrast this with the Serre equations governing gravity-capillary free surface waves, where the equations of motion are given by
\begin{align}
-cu_x+ u u_x  &= -\frac{u_x}{(c-u)^2} + B\mu^2 \lrr{\frac{u}{c-u}}_{xxx}+  \frac{\mu^2}{3h} \lrs{h^3 \lrr{-cu_{xx} + u u_{xx} - u_x^2 }}_x.\label{eq:STnew}
\end{align}
The difference in these systems highlights that, while the same interface displacements are recovered via the relation $B=\nu^2$ for the weakly nonlinear theories discussed in Section~\ref{section:wnl} (since for weakly nonlinear theory both enter only as a correction to the linear dispersion), differences appear in the nonlinear terms of order $O(\mu^2 \epsilon)$.

\begin{figure}
    \centering
    \begin{overpic}[scale=1]{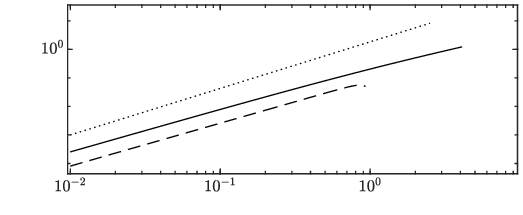}
    \put(0,21){$E$}
    \put(55,0){$A$}
    \end{overpic}
   \caption{Log-log plot of the error $E$ between the KdV and Serre systems for the elevation branch (solid curve) and depression branch (dashed curve) as a function of the amplitude $A$. The dotted curve is included for comparison and shows a line with slope $3/2$.}
    \label{fig:errors}
\end{figure}

Solitary plane waves in the system with $f=0$ are approximated at small amplitudes by the KdV equation~\eqref{eq:KdV}, with an explicit form for the solitary wave given by~\eqref{eq:KDV_sol}. The KdV equation predicts solitary waves of elevation when $\nu<\nu^*=1/\sqrt{3}$, and of depression when $\nu>\nu^*$. We find this is in agreement with the strongly nonlinear solutions, as presented in Figure~\ref{fig:FNsoln}. In panel $(a)$, two branches of stronly nonlinear solitary waves are shown by solid black curves in speed-amplitude parameter space. Here, we choose the amplitude parameter to be the interface perturbation at $x=0$, given by
\begin{align}\label{eq:A}
A=h(0)-1= \eta(0).
\end{align}
We plot one elevation branch with $\nu=0.25$, and one depression branch with $\nu=1$. These branches bifurcate from the long-wave speed $c=1$, and are found in the gap of the linear spectrum in Figure~\ref{fig:disprel}$(a)$. 
The branches are compared with the KdV prediction, shown by the dashed curves in Figure~\ref{fig:FNsoln}. Solutions indicated by crosses and circles in Figure~\ref{fig:FNsoln}$(a)$ are plotted in the respective panels of Figure~\ref{fig:FNsoln}$(b)$--$(e)$. The dashed and solid curves correspond to KdV and Serre solutions respectively. As expected, the KdV theory correctly describes the bifurcation at zero amplitude, but performs quantitatively worse at larger amplitudes. To measure errors we compare the integral value $M$ defined by
\begin{align}
M = \int_{-\infty}^{\infty} \lvert h-1\rvert \, \mathrm{d} x.
\end{align}
We use  $E=\lvert (M - M_\text{KdV})\rvert$ as a measure of the error of the KDV system, where $M_{\text{KDV}}$ refers to the value $M$ computed using the KdV solution. Figure \ref{fig:errors} is a log-log plot of the error $E$ as a function of the amplitude $A$ along the elevation branch (solid curve) and depression branch (dotted curve). For both branches the error is of the order $A^{3/2}$, in agreement with the asymptotic analysis in section \ref{section:wnl}. The relative error, given by $E_r=E/M$, also increases for larger amplitude.
On the elevation branch, solution $(b)$ with $A=0.5$ has $E_r=0.21$, while solution $(c)$ with $A=2.5$ has $E_r=0.50$. Likewise, solution $(d)$ on the depression branch with $A=-0.2$ has $E_r=0.05$, while solution $(e)$ with $A=-0.93$ has $E_r=0.14$. The elevation branch of the Serre system appears to indefinitely increase in amplitude, which the KdV theory also predicts despite the increasing quantitative error. On the other hand, along the depression branch, the KdV solutions can have amplitudes which exceed the depth of the fluid, and are hence no longer physical. The code for the Serre equation becomes stiff for larger amplitude depression solitary waves. Solution $(e)$ is as far as the code can compute solitary waves while satisfying the convergence criterion.

\subsection{Solitary waves for $f\neq0$}\label{section:fneq0}
Next, we consider the case when the Coriolis effect is included. Long waves of small amplitude for this system are approximated by the Ostrovsky equation~\eqref{eq:OS}, which can be written in the same spatial scale $x$ and time scale $t$ as the nonlinear system~\eqref{eq:planewave1}--\eqref{eq:planewave3} as
\begin{equation}
h_t + \lrr{1-\hat{f}\nu \mu^2} h_x +  \frac{3}{2}\lrr{h-1} h_x + \frac{\mu^2}{2} \lrr{\frac{1}{3}-\nu^2} h_{xxx} = \frac{1}{2} \mu^2 \hat{f}^2 \lrr{h-1}
\end{equation}
\begin{figure}[h!]
    \centering
     \begin{overpic}[]{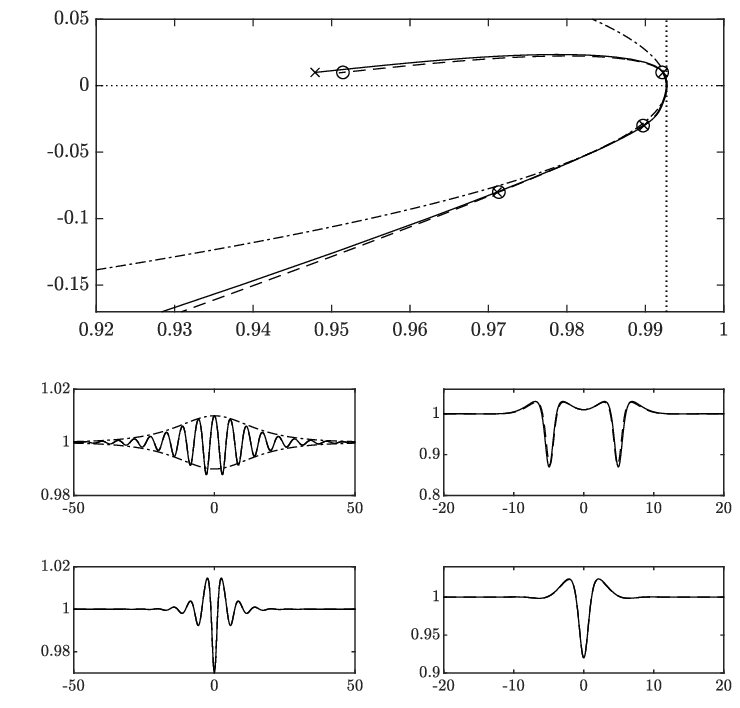}
    \put(42,88){$(c)$}
    \put(91,85.5){$(b)$}
    \put(85,75){$(d)$}
    \put(65,72){$(e)$}
    \put(2,73){$A$}
    \put(55,46){$c$}
    
    \put(2,35){$h$}
    \put(2,11){$h$}
    \put(53,35){$h$}
    \put(53,11){$h$}
    
    \put(0,42){$(b)$}
    \put(0,18){$(d)$}
    \put(52,42){$(c)$}
    \put(52,18){$(e)$}
    
    \put(28,24){$x$}
    \put(28,0){$x$}
    \put(78,24){$x$}
    \put(78,0){$x$}
    
    \end{overpic}
    \caption{Panel $(a)$ shows branches of solitary wavepackets with $\mu=0.2$, $\hat{f}=1$, and $\nu=1$. The solid curves are solutions to Serre equations, the dashed curves are steady  solutions of the corresponding Ostrovsky equation~\eqref{eq:OS}, while the dotted-dashed curves are solutions to the NLS equation~\eqref{eq:NLS}. To four significant figures, the values of the dispersion relation minimum are $c_m=0.9926$ for the Serre equations and $c_m^{\text{o}} = 0.9927$ for the Ostrovsky equation. They are plotted with dotted curves, the difference between the two being indistinguishable in the figure.
     The strongly nonlinear system, the Ostrovsky equation, and the NLS equation all have two branches, one of elevation waves and one of depression waves. The branches bifurcate from an infinitesimal periodic wave train at the dispersion relation minimum. The left hand panel shows a local bifurcation diagram near $A=0$, while the right hand panel shows the depression branch for larger amplitudes. The solutions $(b)$--$(e)$ (represented with crosses for Serre and circles for Ostrovsky) are shown in the remaining panels. In panel $(b)$, the dotted-dashed curve is the NLS wavepacket amplitude.}
    \label{fig:FNsoln2}
\end{figure}
\begin{figure}
    \centering
     \begin{overpic}[]{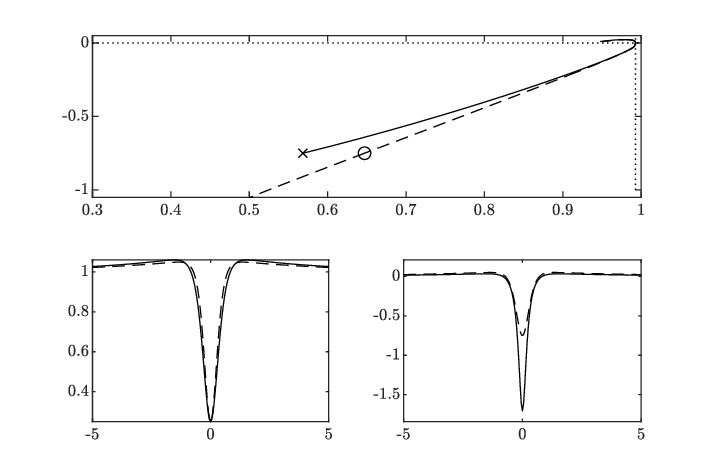}
    \put(5,18){$h$}
    \put(52,18){$u$}
    \put(28,0){$x$}
    \put(74,0){$x$}
    
    \put(50,35){$c$}
    \put(5,50){$A$}
    
    \put(2,60){$(a)$}
    \put(2,28){$(b)$}
    \put(50,28){$(c)$}
    \end{overpic}
    \caption{The solution branches from Figure~\ref{fig:FNsoln2} shown for larger amplitudes. The solid curves are for the Serre system, while the Ostrovsky solutions are shown by dashed curves. The bottom panels show the surface displacement $h$ and the horizontal velocity $u$ for the solution given by the cross and circle for the Serre and Ostrovsky equations, respectively.}
    \label{fig:FNsoln4}
\end{figure}
The far-field conditions are given by equation~\eqref{eq:decay}. Hence, we require the integral term in equation~\eqref{eq:vint} over the whole domain to be zero, i.e.,
\begin{equation}\label{eq:eta_int}
    \int_{-\infty}^{\infty} (h - 1) \, \mathrm{d} x =0.
\end{equation}
Therefore, we require the total volume of water to remain constant. We call the integral in~\eqref{eq:eta_int} the mass of the perturbation. 
When computing solitary waves for non-zero $\hat{f}$, we impose this additional condition, replacing the $x$-momentum equation~\eqref{eq:planewave2} at the first mesh point. We ensure that the converged solution satisfies the $x$-momentum equation at the first mesh point to the same tolerance as the rest of the domain. The form of $v$ for the Ostrovsky equation is given by~\eqref{eq:los}, and hence free-surface perturbations also require zero mean mass. In fact, for the Ostrovsky equation, it can be shown that for periodic and localised solutions, the mass of the perturbation is a conserved quantity and equal to zero.

In Figure~\ref{fig:FNsoln2}$(a)$, we plot solitary wave branches for the Ostrovsky equation (dashed curves) and the nonlinear system~\eqref{eq:planewave1}--\eqref{eq:planewave3} (solid curves) with parameter values $\mu=0.2$, $\hat{f}=1$, and $\nu=1$.
For the odd-viscous Serre equations, the minimum of the phase velocity is given by $c_m\approx 0.9927$ and the corresponding wavenumber is $k_m\approx  1.1046$. For the Ostrovsky system, these values are $c_m^{\text{o}}=0.9926$ and $k_m^{\text{o}}\approx 1.1067$. The dotted-dashed curves are the bifurcation curves for the NLS approximation~\eqref{eq:NLS} of the Serre system, with $k=k_m$. We note that the Ostrovsky solitary waves are not known in explicit form, but are recovered numerically using a  pseudospectral solver akin to the one used to solve the strongly nonlinear system.  There exists one branch of elevation solitary waves and one branch of depression solitary waves bifurcating from the minimum of the linear dispersion relation. The solutions corresponding to the points $(b)$--$(e)$ are shown in their respective panels. For small amplitudes, the solutions are solitary wavepackets, where the carrier wave has a wavenumber approximately equal to the value of $k$ at which the dispersion relation is a minimum. As the amplitude goes to zero, the solution approaches a periodic wave train with this wavenumber. In the small amplitude region, the NLS approximation accurately captures the wavepacket amplitude, as shown in panel $(b)$ of figure \ref{fig:FNsoln2}. Furthermore, the figure shows that the agreement between the Serre and Ostrovsky equations at small amplitudes is superb. For example, using the same measure of relative error as was used to compare the KdV equation and the Serre system, the solutions $(b)-(e)$ have the values $E_r=0.01$, $E_r=0.06$, $E_r=0.02$, and $E_r=0.03$ respectively.
As one follows the elevation branch, two large depressions form, as seen in solution $(b)$. Along the elevation branch, the code fails to converge beyond the solution $(c)$. On the other hand, along the depression branch, the value of $A$ monotonically decreases, with a single large depression at $x=0$, as shown by the solutions $(e)$. Figure~\ref{fig:FNsoln4} shows the depression branch continued into strongly nonlinear regimes, where the deviation between the Ostrovsky and Serre equations increases. In particular, the Ostrovsky equation admits solitary waves with amplitudes larger than the depth of the channel, since this depth is not encoded in equation~\eqref{eq:OS}. Numerical solutions for the Serre system become difficult to compute for solutions past the cross shown in the figure. The solutions begin to form a steep depression about $x=0$, followed by a slow decay to $h=1$ at $x\rightarrow \pm \infty$. This slow decay results in larger computation domains being required to satisfy the condition that the solution be invariant to domain size, yet the region of the solution with a steep depression requires a dense mesh. This combination of requiring both increasing the computational domain and decreasing meshpoint spacing for larger amplitude solutions creates a computational challenge that may better be approached with other numerical methods incorporating variable mesh spacing. The solution shown in figure 10$(b)$ requires a domain of size $L=160$, yet the main depression occurs within $x\in(-1.6,1.6)$.  

%%%%%%%%%%%%%%%%%%%%%%%%%%%%%%%%%%%%%%%%%%%%%%%%

\section{Conclusion}
We have derived nonlinear models describing 3D nonlinear shallow water waves in fluids with nearly vertical odd viscosity, using the results for coarse-grained two-dimensional vortex fluids~\cite{Wiegmann2014}.
Our long-wave isotropic model is an odd-viscous analogue to the Serre equations. 
Odd viscosity enters these equations through typical stress-gradients and, more surprisingly, through terms containing stresses without gradients.
We further simplify the model using a hierarchy of weakly nonlinear unidirectional approximations, leading to KP (as well as the rotation-modified KP), KdV, Ostrovsky, and nonlinear Schr\"{o}dinger equations with odd-viscous contributions.
Through these various reductions we can understand the different manifestation of internal rotation (odd viscosity) versus external rotation (Coriolis forces) on free surface flows. Internal rotations result in surface-tension like effects on the free surface and chiral effects on velocity fields, also allowing for long solitary wave solutions. For example, in the odd-viscous KP equation, odd viscosity acts analogously to a surface tension term and leads to lump solitary waves together with an induced chiral flow. The effects of external rotations alone have been well studied but, for example, preclude long solitary wave solutions enabling instead wavepacket solitary waves.

Odd viscosity is prevalent across many physical systems composed of rotating constituents. These include electrons subject to a magnetic field in two-dimension quantum Hall probes~\cite{avron1995viscosity, berdyugin2019measuring}, and classical chiral active fluids composed of self-rotating particles~\cite{banerjee2017odd, soni2019, han2021fluctuating}. To fix a context, we have focused on a classical vortex fluid, for which odd viscosity can be derived from microscopic models~\cite{Wiegmann2014}.
More generally, many-vortex systems span from quantum states in superfluid helium and cold atomic gases to planetary oceans and atmospheres. In all these cases, we envision exotic solitons on free surfaces, whose specific dynamics remain to be explored.

\appendix

\section{Odd viscosity relative to vortex filament}\label{appendix:OV}

\begin{figure}
\centering
\begin{tikzpicture}
	\node[anchor=south west,inner sep=0] (image) at (0,0) {\includegraphics[scale=.4]{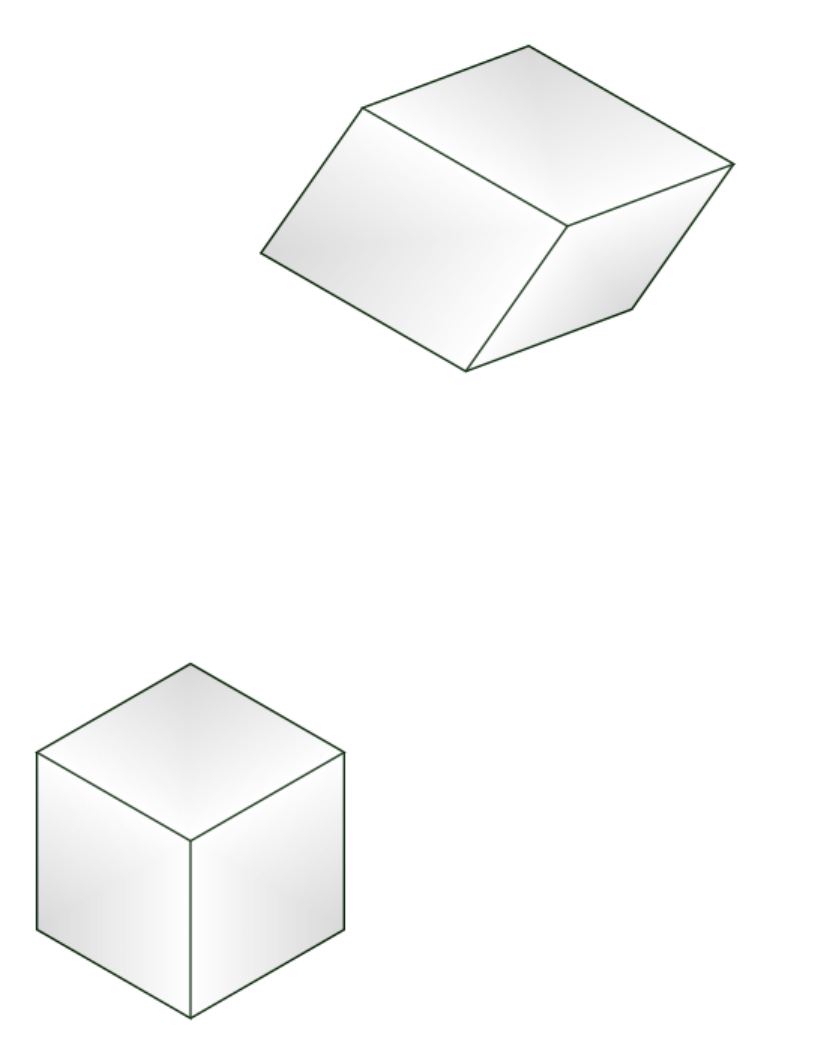}};
	\begin{scope}[x={(image.south east)},y={(image.north west)}]
	\draw [black] plot [smooth] coordinates {(0.25,0.28)  (0.31,0.55) (0.5,0.7)};
	\node [] () at (0.2,0.5) {$\bm{r}(s)$};
	\draw [thick, ->]  (0.55,0.1) -- node[above right,xshift=0.2cm]{$y$} (0.65,0.15);
	\draw [thick, ->]  (0.55,0.1) -- node[below right,yshift=0.15cm,xshift=.2cm]{$x$} (0.65,0.05);
	\draw [thick, ->]  (0.55,0.1) -- node[above,yshift=0.25cm]{$z$} (0.55,0.2);
	\draw [thick, ->]  (0.8,0.55) -- node[above,yshift=0.2cm, xshift=0.1cm]{$z'$} (0.875,0.625);
	\draw [thick, ->]  (0.8,0.55) -- node[below right ,yshift=-0.05cm, xshift=0.3cm]{$x'$} (0.925,0.5);
	\draw [thick, ->]  (0.8,0.55) -- node[above right ,yshift=0.1cm, xshift=0.35cm]{$y'$} (0.925,0.6);
	\end{scope}
\end{tikzpicture}
 \caption{A figure showing a vortex filament and the rotated co-ordinates along the curve. \label{fig:rotation}}
\end{figure}
Consider a vortex filament in the fluid, extending from the bottom wall to the interface at $z=h$. We assume that the filament remains perpendicular with the boundaries at all times. Furthermore, we assume that the vortex filament varies slowly in depth.  Parameterising the vortex filament in arc length $s$, we write its position in  the form
\begin{equation}
\bm{r}(s)=[X(s),Y(s),1-Z(s)]^T.
\end{equation}
Enforcing that the vortex filaments remains perpendicular to the boundaries implies that at the bottom $\bm{r}'=[0,0,1]^T$, while $\bm{r}'$ is equivalent to the unit normal of the surface at $z=h$, where $\bm{r}'$ denotes the $s$ derivative of $\bm{r}$. In nondimensional variables, this can be written as 
\begin{align}
X'&=0, && Y'=0, && Z'=0, && \text{at} \,\, z=0, \\
X'&=-\mu\frac{h_x}{dS}, && Y'=-\mu\frac{h_y}{dS}, && Z'=-\frac{1}{dS},&& \text{at} \,\, z=h(x,y,t).
\end{align}
Here, $dS=\sqrt{1+\mu^2h_x^2+\mu^2h_y^2}$, and $\mu\ll 1$ is the shallow water parameter, introduced in Section~\ref{section:nd}.
Since $dS=1+O(\mu^2)$, we have that at the surface, $X',Y'\sim O(\mu)$ and $Z'\sim O(\mu^2)$. 

We assume that, upon coarse graining, the contribution to the Navier-Stokes equations is an odd viscosity acting in the plane normal to the curve $\bm{r}(s)$. The modified stress tensor $\uu{T}(s)$ along the curve $\bm{r}(s)$ can be recovered by rotating the odd-viscous component of the stress tensor at $s=0$ (that is, at the bottom of the fluid), given by
\begin{align}
\uu{T}(s)\lvert_{s=0} &= 
\mu \uu{K},
\end{align}
with $\uu{K}$ given by equation~\eqref{eq:K} after nondimensionalisation,
to a new set of orthogonal co-ordinates $(\tilde{\bm{x}},\tilde{\bm{y}},\tilde{\bm{z}})$ such that $\tilde{\bm{z}}=\bm{r}'$. The tensor $\uu{K}$ is the odd-viscous stresses that would occur if the vortex filament is purely vertical (i.e. $X'=Y'=Z'=0$ for all $s$), and has been derived for a two-dimensional vortex fluid in \cite{Wiegmann2014}. We write
\begin{align}
\tilde{\bm{x}} &= \begin{bmatrix}1-\alpha_1 \\ \alpha_2 \\ \gamma\beta_3 \end{bmatrix},  &&
\tilde{\bm{y}} = \begin{bmatrix}\beta_1 \\ 1-\beta_2 \\ \beta_3 \end{bmatrix} ,
&&
\tilde{\bm{z}} = \begin{bmatrix}X'\\Y'\\1-Z' \end{bmatrix}.
\end{align}
One can recover the values of $\alpha_i$ and $\beta_i$ given $X',Y'$ and $Z'$ by demanding the three vectors are orthogonal and have magnitude of unity.
There is a degree of freedom to the orthogonal vectors $\tilde{\bm{x}}$ and $\tilde{\bm{y}}$, corresponding to a rotation about the $\tilde{\bm{z}}$ direction, which we keep general with the term $\gamma$ relating the $z$-component of $\tilde{\bm{x}}$ and $\tilde{\bm{y}}$. We desire for this rotation to be small, and furthermore that the vortex filaments do not experience large deflections within the fluid body. 
This is done by imposing that $\alpha_i\ll 1$ and $\beta_i\ll 1$. 
A consequence of this choice is that it introduces a constraint on $\gamma$, which at leading order must be given by $\gamma=X'/Y'$. Hence, we introduce an $O(\mu^2)$ correction, and write $\gamma= X'/Y'+ \mu^2 \hat{\gamma}$.
The rotation matrix $\uu{R}(s)$ for the mapping shown in Fig.~\ref{fig:rotation} is then found to be 
\begin{align}\label{eq:rotmat}
R&= \begin{bmatrix}
1 & 0 & 0 \\
0 & 1 & 0 \\
0 & 0 & 1
\end{bmatrix}
+ \begin{bmatrix}
\frac{X'^2}{2}\mu^2   & (\delta-X'Y')\mu^2 & X'\mu \\
- \delta \mu^2      & \left(-\frac{X'^2}{2}+Z'\right) \mu^2 & Y' \mu \\
- X' \mu              &  -Y'\mu&  Z'\mu^2 
\end{bmatrix}
+ O(\mu^3),
\end{align}
where $\delta$ is given by $\delta=(-\tilde{\gamma}+X'Y'/Z')/(1+X'^2/Y^2)$. One recovers the nondimensional stress tensor along the vortex filament via the equation 
\begin{align}\label{eq:rot1}
\uu{T}(s) = \uu{R}(s) \lrr{\mu\uu{K}} \uu{R}^{-1}(s).
\end{align}
Resolving equation~\eqref{eq:rot1}, we get 
\begin{align}\label{eq:rot2}
\uu{T}(s) = \mu \uu{K} + 
\begin{bmatrix}
O(\mu^3) && O(\mu^3) && -\mu^2 \lrr{X'K_{11}+Y'K_{12}} + O(\mu^3) \\
O(\mu^3) && O(\mu^3) && -\mu^2 \lrr{X'K_{21}+Y' K_{22}}+ O(\mu^3) \\
O(\mu^2) && O(\mu^2) && O(\mu^3) \\
\end{bmatrix}, 
\end{align}
which is at $O(\mu)$ independent of $s$ (i.e., the vortex filament bending).
To describe the equations of motion in full would require knowledge of the bending of the vortex filaments at all points inside the fluid (i.e., values of $\bm{r}'(s)$ must be known for all $(x,y,z)$). However, when deriving the depth-averaged equations (\ref{eq:RSWEcon2}--\ref{eq:RSWE2}) in Section~\ref{section:serre}, it is found that to $O(\mu^3)$ in the model, only the term $\uu{K}$ remains.
 Therefore, the reduced system does not require explicit knowledge of vortex filament bending. 
Instead, we introduce an interpolation function $q(x,y,z,t)$ which takes the value $q=0$ at the bottom and $q=1$ at the surface. The form of the stress tensor is then
\begin{align}\label{eq:ap_stress}
\uu{T}(s) = \mu \uu{K} + 
\begin{bmatrix}
O(\mu^3) && O(\mu^3) && \mu^2 \lrr{h_x K_{11}+h_y K_{12}} + O(\mu^3) \\
O(\mu^3) && O(\mu^3) && \mu^2 \lrr{h_x K_{21}+h_y K_{22}}+ O(\mu^3) \\
O(\mu^2) && O(\mu^2) && O(\mu^3) \\
\end{bmatrix} q. 
\end{align}
The order of each term is considered when nondimensionalising the system in Section~\ref{section:nd}, as in  equations (\ref{eq:ndcont}--\ref{eq:ndmomz}).
Due to the assumed slow variance in depth, it is sufficient to take the lowest order approximation of $q$, given by a linear interpolation of the form
\begin{equation}\label{eq:g}
    q(x,y,z,t) = \frac{z}{h(x,y,t)} + O(\mu).
\end{equation}
We note that the induced stresses do not act along the direction of the vortex filament. This can be shown by first noting that $\uu{T}(0) \bm{r}'(0)=0$. To check it is true for arbitrary $s$, we make use of the rotation matrix~\eqref{eq:rotmat}, to find that
\begin{align}\label{eq:Tfinal}
\uu{T}(s) \bm{r}'(s)&= \left[\uu{R}(s) \uu{T}(0)  \uu{R}^{-1}(s) \right] \left[\uu{R}(s) \bm{r}'(0) \right] =0.
\end{align}
Therefore, at $z=h(x,y,t)$, it is the case that
\begin{align}
T_{ij}n_j=0, && \text{at} \,\, z=h(x,y,t).
\end{align}
The above relation is used to reduce the dynamic boundary condition~\eqref{eq:fsdbc0}.

\section{Derivation of the NLS equation}\label{section:NLS}
Using the method of multiple scales, we introduce slowly varying spatial and time variables:
\begin{align}
X=\e x,  && T=\e t, && \tau = \e^2 t , && \xi = X-c_gT. && \theta = kx - \omega t.
\end{align}
Here, $c_g=d\omega/dk$ is the linear group velocity.
 The unknowns $\bm{\psi}=\lrs{h,u,v}^T$ are sought as a perturbative expansion with small parameter $\e$
\begin{align}
\bm{\psi} &=  \bm{\psi_1} \epsilon + \bm{\psi_2} \e^2 + \cdots.
\end{align}
At leading order, we seek a slowly varying modulated wavepacket propagating with speed $c_g$ in the $x$ direction, with a carrier wave of wavenumber $k$ and frequency $\omega$. This is written as
\begin{align}
\bm{\psi_1} &= \bm{A_{11}}(\xi,\tau) e^{i\theta} + \text{c.c},
\end{align}
where $\bm{A_{11}}$ is a function to be found, and c.c. stands for complex conjugate. The system at $O(\e)$ gives that
\begin{align}\label{eq:leadingorder}
M_1 \bm{A_{11}} &= 0.
\end{align}
where 
\begin{align}
M_n &= \begin{bmatrix}
-in\omega & ink & 0 \\
ink & -in\omega\lrr{1+\frac{1}{3} n^2 \mu^2 k^2 } & -\mu\lrr{\hat{f}- n^2\nu k^2} \\
0 & \mu \lrr{ \hat{f}- n^2 \nu k^2}  & -in\omega 
\end{bmatrix}
\end{align}
To ensure equation~\eqref{eq:leadingorder} has (infinitely many) non-trivial solutions, it must be that $\det(M_1)=0$. This gives rise to the linear dispersion relation $w=w^+$, $w=w^-$, or $w=0$, where
\begin{align}\label{eq:disprelappendix}
%w_{\blue{\pm}} &= \blue{\pm} \lrs{ \frac{\lvert \bm{k} \rvert^2 +\mu^2\lrr{\lrr{\Rey^2\blue{+B}}\lvert\bm{k}\rvert^4 - 2\Rey \hat{f} \lvert \bm{k} \rvert^2  + \hat{f}^2}}{1+\frac{1}{3}\mu^2 \lvert \bm{k}\rvert^2}}^{1/2}, && \text{or} \,\,\,\,\,\, w_{\blue{0}}=0,
\omega_{+} &= + \lrs{ \frac{\lvert \bm{k} \rvert^2  + \mu^2 \lrr{ \Rey \lvert \bm{k} \rvert^2  -\hat{f}}^2}{1+\frac{1}{3}\mu^2 \lvert \bm{k}\rvert^2}}^{1/2}.
\end{align}
We choose the $w^+$ branch, and write $w=w^+$ for the rest of this Appendix. The vector $\bm{A_{11}}$ is chosen to be
\begin{align}
    \bm{A_{11}} &= \begin{bmatrix}
    1 \\ \frac{ \omega}{k} \\ -\frac{i\mu(\hat{f}-\nu k^2)}{k}
    \end{bmatrix} A(\xi,\tau),
\end{align}
where $A$ is the complex amplitude of $h$.
Since $\det(M_1)=0$, a system of the form 
\begin{align}
M_1 \bm{F} &= \bm{G},
\end{align}
for any non-zero vector $\bm{G}$ will have non-trivial solutions $\bm{F}$ if the left eigenvector $\bm{L_1}=[1, \omega/k,i\mu(-\nu k + \hat{f}/k)]^T$  of $M_1$ is orthogonal to the vector $\bm{G}$. This solvability condition will be used at higher order to recover the NLS equation, as shown below. 

We seek a second order solution of the form
\begin{align}
\bm{\psi_2} &= \bm{A_{20}} + \sum_{n=1}^2 \lrs{\bm{A_{2n}}(\xi,\tau) e^{in\theta} + c.c }.
\end{align}
Substituting the above into the system of equations~\eqref{eq:RSWE1}-\eqref{eq:RSWEcon2}, we recover the system 
\begin{align}
M_i \bm{A_{2i}} &= \bm{C_i}, && i=0,1,2,
\end{align}
where
\begin{align}
\bm{C_{0}}&=\begin{bmatrix}
0\\
0\\ 
- 2\mu   \omega \frac{\hat{f}}{k}  \lvert A \rvert^2 \\ 
\end{bmatrix},\\
\bm{C_{1}}&= \begin{bmatrix}
c_g-\frac{ \omega}{k} \\
\frac{ \omega}{k}c_g \lrr{1+\frac{1}{3}\mu^2k^2}  - 1 + \frac{2}{3}\mu^2 \omega^2  - 2 \mu^2 k \nu \lrr{\nu k -\frac{\hat{f}}{k}}   \\
c_g \mu i \lrr{\nu k - \frac{\hat{f}}{k}}- 2\mu \nu i \omega\\ 
\end{bmatrix} A_\xi ,\\
\bm{C_{2}}&= \begin{bmatrix}
-2i\omega \\
\lrr{-\frac{\omega^2}{k}\lrr{1-\frac{5}{3}\mu^2 k^2} - \mu^2 \nu \lrr{\nu k^3 - \hat{f} k } }i  \\ 
\mu \omega \lrr{ 2\nu k - \frac{\hat{f}}{k} }\\ 
\end{bmatrix}A^2.
\end{align}
It can be checked that $\bm{C_1}$ is orthogonal to $\bm{L_1}$. Hence, there are infinitely many solutions $\bm{A_{21}}$, where it is found that the choice has no effect on the NLS equation recovered. We take
\begin{align}
\bm{A_{21}} &= \begin{bmatrix} 0, & \frac{i}{k} \lrr{\frac{\omega}{k} - c_g}, & \mu \lrr{ \nu + \frac{\hat{f}}{k^2}} \end{bmatrix}^T A_\xi.
\end{align}
Solving for $\bm{A_{22}}$, we find 
\begin{align}
\bm{A_{22}} &= \begin{bmatrix} 1 + \mu^2 \lrr{\frac{\hat{f}^2}{k^2} - 2\nu \hat{f}} \\  \frac{\omega}{k}\lrr{2\frac{\omega^2}{k^2} -1  + \mu^2 \lrr{ 2\nu \hat{f} - \frac{3}{2} \frac{\hat{f}^2}{k^2} }  } \\\frac{i\mu}{2k}\lrr{  4 \omega^2 \nu - \hat{f} + \mu^2 \lrr{\hat{f}^2 \nu -  \frac{\hat{f}^3}{k^2} }}  \end{bmatrix} 
\frac{A^2}{2 - 2\frac{\omega^2}{k^2} + \mu^2\lrr{\frac{5}{2} \frac{\hat{f}^2}{k^2} - 4 \nu \hat{f} }}.
\end{align}
Unlike $M_2$, $M_0$ has a zero eigenvalue, with a corresponding left eigenvector $\bm{L_0}=[1,0,0]^T$, which is orthogonal to $\bm{C_0}$. Therefore, there are infinitely many solutions for $\bm{A_{20}}$. We write 
\begin{align}
\bm{A_{20}} &= \begin{bmatrix} P(\xi,\tau) , -\frac{2\omega\mu}{k} \lvert A \rvert^2 , 0 \end{bmatrix}^T,
\end{align}
where the function $P$ is recovered from the oscillation-free terms at the next order.
 We note here the curious fact that the NLS we derive for non-zero $\hat{f}$ is does not reduce to the Coriolis-free NLS equation in the limit as $\hat{f}\rightarrow 0$, such as the one derived in Ref.~\cite{congy2019nonlinear}. The difference stems from second-order contributions to the mean flow. The linear operator for $k=0$, $\omega=0$ in the Coriolis-free case is the zero matrix, while for non-zero $\hat{f}$, it is $M_0$. For $\hat{f}=0$, the vector $\bm{A_{20}}$ would take the form $\bm{A_{20}}=[P,Q,R]$ with $P$, $Q$, and $R$ being functions of $\xi$ and $\tau$ recovered from the oscillation-free terms at $O(\e^3)$. To find solitary wavepackets, which require a minimum in the dispersion relation (see discussion in section~\ref{section:wnl}), we focus on the case of non-zero $\hat{f}$.
 
At $O(\e^3)$, we  seek a solution of the form
\begin{align}
\bm{\psi_3} &= \bm{A_{30}} + \sum_{n=1}^3 \lrs{\bm{A_{3n}}(\xi,\tau) e^{in\theta} + c.c }.
\end{align}
This results in a system of equations given by
\begin{align}
M_i \bm{A_{3i}} &= \bm{D_i}, && i=0,1,2,3,
\end{align}
where 
\begin{align}
    \bm{D_0} &=
    \begin{bmatrix}
     P_\xi\\ - P_\xi+  \lrr{k^2 \nu^2 + \hat{f}\nu - \frac{\omega^2}{k^2}\lrr{1-\frac{1}{3}\mu^2 k^2} - 2c_g \frac{\omega}{k} \lrr{1+\frac{1}{3}\mu^2 k^2} }\lrr{\lvert A\rvert^2}_\xi   \\  \frac{\hat{f}}{k^2} \lrr{c_g k-\omega} \lrr{A^* A_{\xi} - A A_{\xi}^*}
    \end{bmatrix} , \nonumber 
    \\
     \bm{D_1} &=
-\bm{A_{11}} A_{\tau }
+ 
\begin{bmatrix}
-i\omega\\ \frac{2}{3}i k \mu^2 \omega^2 \\0
\end{bmatrix}
P A +
\begin{bmatrix}
- \frac{i}{k} \lrr{\frac{\omega}{k}-c_g} \\
i \lrr{3k\nu^2 + \frac{\hat{f}\nu}{k} + \frac{1}{3k}\mu^2 \omega^2 - \frac{4 }{3} c_g \mu^2 \omega  + c_g \gamma \frac{\omega}{k^2} -\frac{1}{k} c_g^2 \gamma }\\
\frac{\nu \omega}{k}  + c_g \frac{\lambda}{k^2}
\end{bmatrix}
A_{\xi\xi} \nonumber 
\\
&+ 
\begin{bmatrix}
3i\omega - 2i\omega \delta \\
i\lrr{\frac{\omega^2}{k}\lrr{3-\frac{1}{3} \mu^2 k^2} - \hat{f}k\nu - k^3\nu^2 + \lrr{-\frac{\omega^2}{k^2}\lrr{1+\frac{1}{3}\mu^2 k^2} + 2k^3 \nu^2 + \hat{f}k\nu             }\delta } \\
\omega\lrr{\frac{3\hat{f}}{k} - 2k\nu} + \omega \lrr{k\nu - \frac{2\hat{f}}{k}} \delta
\end{bmatrix}
\lvert A \rvert^2 A, \nonumber 
\end{align}
where for simplicity we define the coefficients
\begin{align}
    \gamma &= 1+\frac{1}{3} \mu^2 k^2, &&
    \delta =  \frac{\lrs{1-2\nu \hat{f} + \frac{\hat{f}^2}{k^2} }  }{2 - 4 \nu \hat{f} + \frac{5}{2} \frac{\hat{f}^2}{k^2}-2 \frac{\omega^2}{k^2}}. 
\end{align}
The solvability condition requires both that  $\bm{D_0}$ is orthogonal to $\bm{L_0}$, resulting in $P_\xi=0$,  and that $\bm{D_1}$ is orthogonal to $\bm{L_1}$, resulting in the celebrated NLS equation
\begin{align}\label{eq:NLS_ap}
iA_\tau + \frac{1}{2} \frac{d^2\omega}{dk^2} A_{\xi\xi} = \beta \lvert A \rvert^2 A,
\end{align}
where 
\begin{align} \label{eq:beta} 
\beta &= - \omega\frac{ 2 \mu^4   \hat{f}^2   (\hat{f} - k^2  \nu )(\hat{f} - 2k^2  \nu ) + \mu^2  \hat{f}  (k^2+2  \omega^2) (3  \hat{f}-4  k^2  \nu) +  k^4 + 4  k^2  \omega^2 - 8  \omega^4}{\lrs{\mu^2 \hat{f}(5 \hat{f} - 8 k^2\nu )+ 4k^2 - 4\omega^2}\lrs{\mu^2 (\hat{f} -  k^2 \nu )^2 + k^2}} .
\end{align}

%\section*{Acknowledgments}

\bibliography{slack.bib}

\end{document}